\documentclass[11pt,aps,showpacs]{revtex4-1}
\usepackage[T1]{fontenc}
\usepackage[utf8]{inputenc}

\usepackage{graphicx}

\usepackage{dcolumn}
\usepackage{bm}
\usepackage{mathrsfs}
\usepackage{amsfonts}
\usepackage{amssymb}
\usepackage{amsmath}

\def\nslash{n\!\!\!\slash}
\def\bnslash{\bar n\!\!\!\slash}

\usepackage{slashed}

\newcommand{\nn}{\nonumber} 
\newcommand{\bn}{{\bar n}}
\newcommand{\mcdot}{\!\cdot\!}

\newcommand{\scetii}{\text{SCET}_\text{II}}
\newcommand{\LQCD}{\Lambda_{\rm QCD}}
\newcommand{\arccosh}{\mathrm{arccosh}}

\newcommand{\paren}[1]{\left(#1\right)}
\newcommand{\oneov}[1]{\frac{1}{#1}}
\newcommand{\req}[1]{Eq.\,\eqref{#1}}
\newcommand{\beq}{\begin{equation}}
\newcommand{\eeq}{\end{equation}}

\newcommand{\dg}{\dagger}

\begin{document}

\title{Rapidity regulators in the semi-inclusive deep inelastic scattering and Drell-Yan processes}
\author{Sean Fleming$^1$ and Ou Z. Labun$^{1,2}$}
\affiliation{$^1$\,Physics Department, University of Arizona, Tucson, Arizona, 85721, USA \\
$^2$\,Institut de Physique Nucl\'{e}aire, IN2P3-CNRS, Universit\'{e} Paris-Sud, F-91406, Orsay Cedex, France}

\begin{abstract}
We study the semi-inclusive limit of the deep inelastic scattering and Drell-Yan (DY)  processes in soft collinear effective theory.  
In this regime so-called threshold logarithms must be resummed to render perturbation theory well behaved. Part of this 
resummation occurs via the Dokshitzer, Gribov, Lipatov, Altarelli, Parisi (DGLAP) equation, which at threshold contains a large logarithm 
that calls into question the convergence of the anomalous dimension. We demonstrate here that the problematic logarithm is 
related to rapidity divergences, and by introducing a rapidity regulator can be tamed. We show that resumming the rapidity 
logarithms allows us to reproduce the standard DGLAP running at threshold as long as a set of potentially large nonperturbative 
logarithms are absorbed into the definition of the parton distribution function (PDF). These terms could, in turn, explain the steep falloff
of the PDF in the end point. We then go on to show that the resummation of rapidity divergences does not change the standard
threshold resummation in DY, nor do our results depend on the rapidity regulator we choose to use.

\end{abstract}


\maketitle

\section{Introduction}\label{sec:intro}

Lepton pair production in hadron-hadron collisions, known as
the Drell-Yan (DY) process, helped establish the parton model as a valid leading-order
description of high energy QCD interactions.  At present the DY
process is still of great interest as it provides a test bed for other final states, such as
the Higgs boson or beyond-the-standard-model particles, which are similarly produced
in the collision of high energy partons \cite{drell70}.  

Of particular theoretical interest is the so-called threshold region, 
where the invariant mass of the lepton pair approaches the center-of-mass energy of the collision. In this regime
large Sudakov logarithms must be resummed~\cite{sterman1987,catani1989, magnea1991,korchemsky1993}.
Similar, but on less rigorous footing is the need for partonic resummation. In this case one is not in the true end point
region, but rather in the region where the invariant mass of the colliding partons is just above the threshold
for the production of the final state. It is argued~\cite{appell1988,catani1998} that the sharp falloff of parton 
luminosity at large $x$ enhances the partonic threshold region, and thus requires resummation. A quantitative study of this question was carried out
in the context of soft collinear effective theory (SCET)~\cite{bauer02,bauer00,bauer01,Bauer:2002nz} in Ref.~\cite{becher07}, which concludes among other things
that ``the dynamical enhancement of the threshold contributions remains effective down to moderate values
$\tau \approx 0.2$,'' where $\tau =1$ represents the true end point.

In the threshold region the large Sudakov logarithms which need to be resummed have a simple form in Mellin moment space, where 
leading terms appear in perturbation theory as double logarithms $\alpha_{s}^{n}\ln^{2n}(N)$, where $N$ is the Mellin moment. 
The threshold region corresponds to the limit of large $N$, so clearly the presence of these types of terms poses problems for a naive perturbative
expansion and calls for resummation. Part of this resummation occurs when the parton distribution function (PDF) is evolved 
using the Dokshitzer, Gribov, Lipatov, Altarelli, Parisi (DGLAP)~\cite{gribov1972,altarelli1977,dokshitzer1977} equation, which in the threshold region becomes particularly simple. In 
Mellin moment space the anomalous dimension for the nonsinglet quark-to-quark PDF has the form~\cite{moch2004}
\begin{equation}
\label{mellmomanom}
\gamma_{ns}^{(n)} = -\bigg(\frac{\alpha_{s}(\mu)}{4 \pi}\bigg)^{n+1}\bigg[ A_{n}\log(\bar{N}) - B_{n}\bigg] + {\mathcal O}\bigg(\frac{\ln(N)}{N},\frac{1}{N}\bigg)\,,
\end{equation}
where $\bar{N} = N e^{\gamma_{E}}$, $\gamma_{E}$ being the Euler-Mascheroni constant. At order $n=0$, for example, $A_{0} = 16/3\approx 5.3$ and $B_{0}= 4$. 
What is peculiar about this result is that while $A_{n}$ and $B_{n}$ are numbers of the same order, there is the large logarithm of $N$ enhancing the $A_{n}$ term. 
From an effective field theory (EFT) point of view the large logarithm is problematic because a consistent power counting in the threshold region should never encounter such enhanced terms. 

This issue was addressed in a previous paper in which we revisited deeply inelastic scattering (DIS) in the threshold (or end point) region, 
where Bjorken-$x$ approaches its end point value of one~\cite{Fleming:2012kb}. In that work we use SCET to show that the PDF in the threshold region can be
expressed as the product of a collinear factor and a soft function. Since both the collinear and soft degrees of freedom in the end point have an invariant mass 
of order the hadronic scale such a separation necessitates the introduction of a rapidity regulator to keep the two modes separate. We use the rapidity regulator
of Refs.~\cite{Chiu:2011qc,Chiu:2012ir}. This tool allows us to reorganize the perturbative expansion of the
anomalous dimension for the nonsinglet quark-to-quark PDF in the threshold region. We find the leading-order anomalous dimension in Mellin moment space to be
\begin{equation}
\gamma_{ns}^{(0)} = -\bigg(\frac{\alpha_{s}(\mu)}{4 \pi}\bigg)^{n+1}\bigg[ A_{0}\ln \bigg(\frac{\nu_{c}\nu_{s}}{Q^{2}/\bar{N}}\bigg) - B_{0}\bigg] \,,
\end{equation}
where $\nu_{c}\approx Q$ is the collinear rapidity scale, and $\nu_{s} \approx Q/\bar{N}$ is the soft rapidity scale. The rapidity scales are set by 
minimizing logarithms in the collinear and soft anomalous dimensions, and result in a $\gamma_{ns}^{(0)} $ free of a logarithmic enhancement. 
Now both terms in the anomalous dimension are of ``natural'' size, ${\mathcal O}(1)$.

Unfortunately, there is a downside to separating modes in rapidity: the PDF now depends on logarithms of the ratio of $\nu_{c}$ to $\nu_{s}$. In principle these 
logarithms can be resummed using a rapidity renormalization group equation (rRGE); however the anomalous dimension in the rRGE is not infrared safe. As 
a result the running in rapidity can not be reliably calculated and must be included in the function chosen to model the PDF at the hadronic scale. This does not mean
that we cannot use our rapidity separated PDF as the definition of the PDF in the end point: we can as long as we let the scale $\nu_{s}$ approach $Q$ as we move
away from threshold. This can be achieved by introducing a rapidity profile function~\cite{abbate2011}.

In our previous work we showed that one can introduce a rapidity separated PDF in the end point of DIS that has all the properties that a PDF should have, and that DIS
in the end point using our approach factors in the same way as DIS factors in the region away from the end point.  This approach, however, must also reproduce the well-known result in DY that threshold resummation  is expressed as a convolution of perturbatively 
resummed logarithms with the same PDF as appears in DIS. The aim of this paper is to  show that this is indeed the case. Furthermore, we investigate an alternative 
rapidity regulator, the delta regulator, and show that our results are rapidity regulator independent to the order we are working.

We begin in Sec. \ref{sec:sec2} by reviewing our calculation of the DIS soft and collinear functions using the $\eta$ regulator.  In Sec. \ref{sec:sec3}, we calculate the soft and collinear functions for DY using the $\eta$ regulator and resum the end point logarithms using the rapidity renormalization group.  In Sec. \ref{sec:sec4}, we repeat the calculations for both DIS and DY using the delta regulator and compare the results to those from using the $\eta$ regulator.  For completeness, in Appendix \ref{app:jetfndeltareg} we calculate the jet function (for DIS) using the delta regulator, which has not previously appeared in the literature.  We explore the difference in the structure of the zero-bin subtraction between DY and DIS in Appendix \ref{app:zerobin}.

\section{DIS at the end point with the Rapidity Regulator}\label{sec:sec2}
In this section, we review the SCET factorization and resummation results for DIS in the end point regime which we studied in Ref.\,\cite{Fleming:2012kb}. At the end of this section we remark on aspects of our results that were not addressed in our previous work, and compare to previous work~\cite{Becher:2006mr}.

The DIS process is when a high energy electron with momentum $k$ strikes a proton with momentum $p$ and produces to a final hadronic state $X(p_X)$ and a scattered electron. We denote the final state electron momentum as $k'$, and the square of the momentum transfer is $q^2=(k-k')^2$. We define $Q^2\equiv -q^2$, and $x=\dfrac{Q^2}{2p\cdot q}$. With this notation, we follow Ref.\,\cite{Fleming:2012kb} and write the differential cross section as
\begin{equation}\label{eq:1'}
d\sigma=\frac{d^3\vec k}{2|\vec k'|(2\pi)^3}\frac{\pi e^4}{SQ^4}L_{\mu\nu}(k,k')W^{\mu\nu}(p,q),
\end{equation}
where $s=(p+k)^2$ is the invariant mass square of the collision, and the lepton tensor is
\begin{equation}\label{eq:1''}
L_{\mu\nu}=2(k_\mu k'_\nu+k_\nu k_\mu'-k\cdot k' g_{\mu\nu})\,.
\end{equation}
$W_{\mu\nu}$ is the DIS hadronic tensor, which at large $x$ is the subject of our analysis.  

In this section, we first determine the kinematics and power-counting specific to the end point.  Then we match QCD onto SCET${}_{\text{I}}$.  Next at an intermediate scale of order the invariant mass of the final state, we match the SCET${}_{\text{I}}$ onto SCET${}_{\text{II}}$. Using the rapidity regulator introduced in Refs \cite{Chiu:2011qc,Chiu:2012ir}, we explicitly calculate both the collinear and the soft functions to one loop in the SCET${}_{\text{II}}$.

\subsection{Kinematics}\label{sec:sec2.1}
There are a number of different approaches in the literature \cite{Manohar:2003vb,Chay:2004zn,Chay:2013zya} that describe how momentum components separate and scale in the $x\sim 1$ regime. In this article, we choose the notations in Ref.\,\cite{Manohar:2003vb}.  We define light-cone unit vectors $n^\mu=(1,0,0,-1)$ and $\bar n^\mu=(1,0,0,1)$, which allows us to decompose the proton momenta $p^\mu=\frac{n^\mu}{2}\bar n\cdot p+\frac{\bar n^\mu}{2}n\cdot p+p_\perp^\mu$, in which $p^+=n\cdot p$ and $p^-=\bar n\cdot p$. In the target rest frame, $p=(p^+,p^-,p_\perp)=(M_p,M_p,0)$, and $Q^2=-q^2=-q^+q^-$. The direction of the incoming electron fixes the z-axis, and in the target rest frame, $q^-\gg q^+$.  In this limit, Bjorken $x$ simplifies to
\begin{equation}\label{bjorkenx}
x=\frac{Q^2}{2p\cdot q}=-\frac{q^+q^-}{p^+q^- + p^-q^+}\simeq -\frac{q^+}{p^+}\,.
\end{equation}
We can express all momenta in terms of $x$, $M_p$ and $Q$ in the target rest frame, and then boost them along the z-axis into the Breit frame,
\begin{eqnarray}
q&=&\left(-x M_p,\frac{Q^2}{xM_p}, 0 \right) \xrightarrow{\text{boost}} \left(-Q,Q,0\right)\nonumber \\
p&=&\left(M_p, M_p, 0\right) \xrightarrow{\text{boost}} \left(\frac{Q}{x}, \frac{xM_p^2}{Q}, 0\right)\nonumber \\
p_X&=&p+q=\left(M_p(1-x), q^-, 0\right)\nonumber 
\xrightarrow{\text{boost}} \left(\frac{Q(1-x)}{x}, Q,  0\right)\,,
\end{eqnarray}
where $p_X$ is the (total) final state momentum.
In the large-$x$ limit, the large component of the incoming proton is $p^+=\frac{Q}{x}=Q+l^+$, in which $l^+=Q\frac{1-x}{x}$ is a rapidity scale lying between the collinear momentum scale $Q$ and soft momentum scale $\Lambda_{\text{QCD}}$.  The rapidity scale, as we see later, separates soft and collinear modes and gives rise to logarithms of $\nu_s$ and $\nu_c$. Correspondingly, we have naturally separated momenta,
\begin{itemize}
\item hard modes with $q\sim \left(-Q, Q, 0 \right)$ and invariant mass $q^2\sim Q^2$ at the hard collision scale;
\item final state jet hard-collinear modes with $p_X\sim \left(Q\paren{\frac{1-x}{x}}, Q, Q\sqrt{\frac{1-x}{x}}\right)\sim \left(l^+, Q,\sqrt{Ql^+}\right)$ and invariant mass $p_X^2\sim Ql^+ \gg \Lambda_{\text{QCD}}^2$ at the hard-collinear scale;
\item $n$-collinear modes with $p_c\sim \left(Q, \frac{\Lambda_{QCD}^2}{Q}, \LQCD\right)$ and invariant mass $M_p^2\sim \Lambda_{\text{QCD}}^2$ at the soft scale;
\item soft modes with $p_s\sim (\LQCD,\LQCD,\LQCD)$ at the soft scale.
\end{itemize}
We first integrate out the hard degrees of freedom in QCD at the scale $Q^2$ by matching onto $\text{SCET}_{\text{I}}$ with off-shellness $Ql^+$. We then integrate out hard-collinear degrees of freedom at $Ql^+$ by matching onto  $\text{SCET}_{\text{II}}$ with off-shellness $\LQCD^2$.  In the case where the final state momentum $p_X^+$ is of order $Q\paren{\frac{1-x}{x}}\sim l^+\gtrsim \Lambda_{\text{QCD}} \ll Q$, the process is semi-inclusive in character. If on the other hand $l^+\sim \frac{\Lambda_{\text{QCD}}^2}{Q}$, the collision would be exclusive, and we would be unable to factor the hadronic tensor.

\subsection{Factorization}\label{sec:sec2.2}
In Eq.\eqref{eq:1'}, the DIS hadronic tensor is the matrix element of the time-ordered product of  two QCD currents $J^\mu(x)=\bar \psi(x)\gamma^\mu \psi(x)$ between external in- and out-proton states,
\begin{eqnarray}\label{eq:3}
W^{\mu\nu}(p,q)&=&\frac12 \sum_\sigma \int d^4x e^{iq\cdot x} 
\langle h(p,\sigma) \vert J^\mu(x) J^\nu(0) \vert h(p,\sigma) \rangle,
\end{eqnarray}
where $\sigma$ is the spin of the proton.  Matching QCD onto SCET is carried out at the scale $\mu_q\sim Q$, and the SCET current is
\begin{eqnarray}\label{QCDSCETcurrent}
J^\mu(x)&\to& \sum_{w_1,w_2}C(w_1,w_2;\mu,\mu_q)
\left(e^{-\frac i2 w_1n\cdot x}e^{\frac i2 w_2\bar n\cdot x}\bar \chi_{\bar n,w_2}\gamma_\perp^\mu \chi_{n,w_1}+\text{h.c.}\right)\,,
\end{eqnarray}
where $\bar\chi_{\bar n,w_2},\chi_{n,w_1}$ are SCET fields.  Correspondingly, the hadronic tensor in $\text{SCET}_{\text{I}}$ is
\begin{eqnarray}
W_{\text{eff}}^{\mu\nu}&=&\sum_{w_1,w_2,w_1',w_2'}C^*(w_1,w_2;\mu_q,\mu)C(w_1',w_2';\mu_q,\mu) \int\frac{d^4x}{4\pi}e^{-\frac i2(Q-w_1)n\cdot x}e^{\frac i2 (Q-w_2)\bar n\cdot x}\nonumber \\
&&\times \oneov{2} \sum_\sigma \langle h_n(p,\sigma)\vert \bar T[\bar \chi_{n,w_1}\gamma_\perp^\mu \chi_{\bar n,w_2}(x)] T[\bar \chi_{\bar n,w_2'}\gamma_\perp^\nu \chi_{n,w_1'}(0)]\vert h_n(p,\sigma)\rangle \nn \\
&=& \frac{- g_\perp^{\mu\nu}}{2} N_c 
 \sum_{\omega'_1, \omega'_2} C^*(Q ,Q;\mu_q, \mu ) C(\omega'_1,\omega'_2;\mu_q, \mu)\nn\\
&&\times \int \frac{d^4 x}{4 \pi}\frac{1}{2} \sum_\sigma  \langle{h_{n}(p,\sigma)}|  \bar{\chi}_{n, Q}(x)\frac{\bnslash}{2} \chi_{n, \omega'_1}(0)|{h_{n}(p,\sigma)}\rangle \nn\\
&&\times \langle{0} |\frac{\nslash}{2}\chi_{\bn, Q}(x) \bar{\chi}_{\bn, \omega'_2}(0) |0\rangle  \frac{1}{N_c} \langle{0} |\text{Tr}\bigg(
\bar{T}\bigg[ Y^\dagger_{n}(x) \tilde{Y}_{\bn}(x)\bigg]
T\bigg[ \tilde{Y}^\dagger_{\bn}(0) Y_{n}(0) \bigg]
\bigg)|0\rangle \,, \label{eq:05}
\end{eqnarray}
where $T$ and $\bar T$ denote time ordering and antitime ordering operations of the soft gluon fields ${Y}_{\bn}$ and $Y_{n}$ respectively. The two collinear sectors and one usoft sector are decoupled by the BPS phase redefinition in Ref.\,\cite{bauer02}.

In order to match Eq.\,\eqref{eq:05} onto $\text{SCET}_{\text{II}}$, it is convenient to introduce a jet function as in Ref.\,\cite{Fleming:2007xt},
\begin{eqnarray}
&&\langle{0}|\frac{\bnslash}{2}\chi_{\bn, \omega_2}(x) \bar{\chi}_{\bn, \omega'_2}(0)|0 \rangle
\equiv  Q \delta(\bn\cdot x) \delta^{(2)}(x_{\perp}) \int dr \, e^{-\frac{i}{2}r n \cdot x} J_{\bn}(r;\mu) \,,\label{eq:06}
\end{eqnarray}
which characterizes the final state with $p_X^2\sim Ql^+$. The final state is integrated out at the scale $\mu_c\sim \sqrt{Ql^+}$ and $J_{\bar n}(r;\mu)$ becomes a matching coefficient in $\text{SCET}_{\text{II}}$. 

We define a soft function in $\text{SCET}_{\text{I}}$ as in Ref.\,\cite{bauer03},
\begin{eqnarray}
&&\frac{1}{N_c} \langle{0} |\text{Tr}\bigg(
\bar{T}\bigg[ Y^\dagger_{n}(n\cdot x) \tilde{Y}_{\bn}(n\cdot x) \bigg]
\text{T}\bigg[\tilde{Y}_{\bn}^\dg(0) Y_{n}(0) \bigg]
\bigg)|0\rangle \equiv  \int d\ell \, e^{-\frac{i}{2}\ell n\cdot x} S^{(DIS)}(\ell;\mu) \,,\label{eq:07}
\end{eqnarray}
which describes usoft gluon emission throughout the interaction, from the initial to final state.  The Wilson lines are defined as
\begin{align}
Y_n(x)&=P\exp\paren{ig\int_{-\infty}^xds\,\,\, n\cdot A_s(s_n)}\,,\nn\\
\tilde Y_{\bar n}^\dg (x)&=P\exp\paren{ig\int_x^\infty ds\,\,\, \bar n\cdot A_s(s_{\bar n})}. \label{eq:13'}
\end{align}
The usoft gluons in $\text{SCET}_{\text{I}}$ with off-shellness $p_{us}^2\sim \Lambda_{\text{QCD}}^2$ become soft gluons of $\text{SCET}_{\text{II}}$, so  Eq.\eqref{eq:07} retains its form in matching $\text{SCET}_{\text{I}}$ to $\text{SCET}_{\text{II}}$.

Using label momentum conservation, which is just momentum conservation at fixed (large) $Q$, we simplify the collinear matrix element in the $n$-collinear direction,
\begin{eqnarray}
 \langle{h_{n}(p,\sigma)} |\bar{\chi}_{n,  Q}(x)\frac{\bnslash}{2} \chi_{n, \omega'_1}(0)|{h_{n}(p,\sigma)}\rangle 
=\delta_{ Q,\omega'_1} \,\langle{h_{n}(p,\sigma)}| \bar{\chi}_{n}(x)\frac{\bnslash}{2} \delta_{\bar{\cal P}, 2  Q}  \chi_{n}(0) |{h_{n}(p,\sigma)}\rangle \,.\label{eq:08}
\end{eqnarray}
We then define an $n$-direction collinear sector as the $n$-collinear function and match it onto $\text{SCET}_{\text{II}}$.  We insert an explicit Kronecker delta to ensure the large momentum of the proton $\tilde p\cdot\bar n$ is $Q$ at large $x$,
\begin{eqnarray}
{\cal C}_{n}( Q-k;\mu)  
&=& \int \frac{d\, n\mcdot x }{4 \pi}\, e^{\frac{i}{2}kn\cdot x} \frac{1}{2} \sum_\sigma  
\delta_{\bn\cdot\tilde p,  Q}\,
 \langle{h_{n}(p,\sigma)}| \bar{\chi}_{n}(n\mcdot x)\frac{\bnslash}{2} \delta_{{\cal \bar P}, 2  Q} \chi_{n}(0) |{h_{n}(p,\sigma)}\rangle \nn \\
&=&\frac{1}{2} \sum_\sigma  
\delta_{\bn\cdot\tilde p,  Q}\, \langle{h_{n}(p,\sigma)}| \bar{\chi}_{n}(0)\frac{\bnslash}{2} \delta_{\bar{\cal P}, 2  Q} \delta(i\bn\cdot\partial -k) \chi_{n}(0) |{h_{n}(p,\sigma)}\rangle\,,\label{eq:09}
\end{eqnarray}
where ${\cal \bar P}= {\bar n} \cdot(\cal P+\cal P^+)$ and $k\sim \LQCD$ is the residual momentum lying in the $\text{SCET}_{\text{II}}$ soft region.  Label momentum conservation then forces $w_1'=Q$, meaning that the large momenta of the incoming and outgoing protons are both equal to $Q$. 

In the $\text{SCET}_{\text{II}}$ soft and collinear fields have the same off-shellness $p^2\sim \Lambda_{\text{QCD}}^2$. An arbitrary separation between these soft and collinear modes may lead to rapidity divergences \cite{Chiu:2011qc,Chiu:2012ir}, which we regulate by a Lorentz invariant $\eta$ regulator with a dimensionful scale $\nu$. Since the matching procedure shows that the final state jet function is decoupled from the initial state $n$-collinear function, we can express the $n$-collinear function as $C_n(Q-k;\mu)\to C_n(Q-k;\mu,\nu)$
and the soft function as $S(l,\mu)\to S(l;\mu,\nu)$.  Combining Eqs.\eqref{eq:05}, \eqref{eq:06}, \eqref{eq:08} and \eqref{eq:09}, we arrive at the $\text{SCET}_{\text{II}}$ factorized DIS hadronic tensor,
\begin{eqnarray}
\label{hadtenSCET2}
W^{\mu \nu}_{\rm eff} = -g_\perp^{\mu\nu}H( Q ;\mu_q, \mu_c)   
 \int d \ell \, J_{\bn}(\ell;\mu_c,\mu) f^{ns}_q(Q\paren{\frac{1-x}{x}}+\ell;\mu), \label{eq:010}
\end{eqnarray}
with
\begin{equation}\label{eq:011}
f_q^{ns}\paren{\ell;\mu}=\delta_{\tilde n\cdot\tilde p,Q}\mathcal{Z}_n(\mu,\nu)S^{(DIS)}(\ell;\mu,\nu)
\end{equation}
and
\begin{equation}\label{eq:012}
\mathcal{Z}_n(\mu,\nu)=C_n(Q-k;\mu,\nu)\delta(k)\delta_{\bar n\cdot\tilde p,Q}\,.
\end{equation}

\subsection{Renormalization and resummation with rapidity}\label{sec:sec2.3 }
In this section we study the collinear and soft functions using the $\eta$ regulator from Refs.\,\cite{Chiu:2011qc, Chiu:2012ir}. The rapidity logarithms in the collinear and soft functions are regulated by a modification of the momentum space Wilson lines as follows,
\begin{eqnarray}
W_n=&\sum_{\rm perms}\exp\left[-\frac{gw^2}{\bn\cdot \mathcal{P}}\frac{|\bn\cdot\mathcal{P}|^{-\eta}}{\nu^{-\eta}}\bn\cdot A_n\right]\,, \nn \\
S_n=&\sum_{\rm perms}\exp\left[-\frac{gw}{\bn\cdot \mathcal{P}}\frac{|2\mathcal{P}_3|^{-\eta}}{\nu^{-\eta}}n\cdot A_s\right]\,, \label{Wsofteta}
\end{eqnarray}
where $\nu$ is a rapidity scale and $w$ is analogous to a coupling constant, which is used to derive the rapidity renormalization group equation.  We take $\eta\to0$ at the end.

\subsubsection{Collinear function to $\mathcal{O}(\alpha_s)$ for DIS}
\begin{figure}
\centering
\includegraphics[scale=0.7]{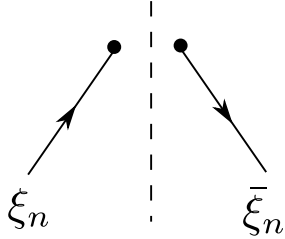}
\caption{$\mathcal{O}(\alpha_s^0)$ Feynman diagram for the $n$-collinear function.}
\label{fig:1}
\end{figure}

The $n$-collinear function in \req{eq:09} has the tree-level Feynman diagram shown in Fig.\,\ref{fig:1}.
We consider the explicit calculation of this diagram using external parton states, and find the $\mathcal{O}(\alpha_s^0)$ result
\begin{equation}\label{eq:12'}
C_n^{(0)}(Q-k)=\delta_{\bar n\cdot \tilde p,Q}\delta(\bar n\cdot p_r-k)m_0\,,
\end{equation}
where $\bar n\cdot \tilde p$ is the $\mathcal{O}(1)$ quark label momentum at the hard scale $Q$, $p_r$ is the quark residual momentum at the soft scale and
\begin{equation}\label{eq:12''}
m_0=\frac12\sum_\sigma\bar\xi_n^\sigma\frac{\bar n \!\!\!\!\!\not\,\,\,\,}{2}\xi_n^\sigma\,,
\end{equation}
where $\xi_n^\sigma$ is the SCET quark spinor in the $n$ direction with spin $\sigma$.

\begin{figure}
\centering
\includegraphics[scale=0.7]{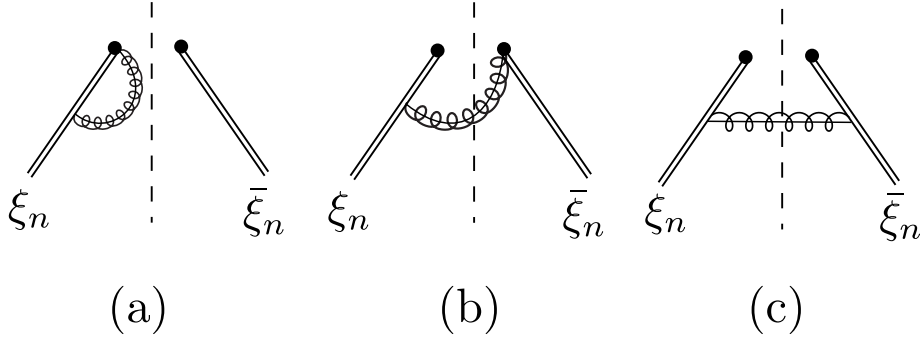}
\caption{The $\mathcal{O}(\alpha_s)$ Feynman diagram for the $n$-collinear function (a) is the virtual contribution; (b) and (c) are the real contribution.}
\label{fig:2}
\end{figure}

The $\mathcal{O}(\alpha_s)$  $n$-collinear function Feynman diagrams are shown in Fig.\,\ref{fig:2}.
Figure.\,\ref{fig:2}(a) shows the virtual contribution, while Figs. \ref{fig:2}(b) and (c) show the real contribution.  We omit the mirror images of Figs.\,\ref{fig:2}(a) and \ref{fig:2}(b).
With the rapidity regulated collinear Wilson lines, we obtain the naive result corresponding to the diagram in Fig.~\ref{fig:2}(a),
\begin{align}
i {\tilde m_a^n}&=(im_0)(2g^2C_F)\delta_{\bar n\cdot p,Q}\delta(l^-)\mu^{2\epsilon}\nu^\eta
\int\frac{d^Dq}{(2\pi)^D}\frac{|\bar n\cdot q|^{-\eta}}{\bar n\cdot q}\frac{\bar n\cdot(p-q)}{(p-q)^2+i\epsilon}\oneov{q^2-m_g^2+i\epsilon}   \label{eq:35}
\end{align}
in $D=4-2\epsilon$ dimensions. The Kronecker delta sets the large component of the external quark momentum to $Q$.  The integral in \req{eq:35} overlaps with a region of soft momenta that must be subtracted to avoid double counting, the so-called zero bin which was first discussed in Ref.\,\cite{Manohar2007} and then improved in Ref.\,\cite{Idilbi:2007yi}.  Taking the limit $\bar n\cdot q\ll \bar n\cdot p$ in the collinear gluon loop gives the overlap region, and the zero-bin subtraction for this diagram is
\begin{align}
im_a^{n\phi}&=im_0(2g^2C_F)\delta_{\bar n\cdot p,Q}\delta(l^-)\mu^{2\epsilon}\nu^\eta
\int\frac{d^Dq}{(2\pi)^D}\frac{|\bar n\cdot q|^{-\eta}}{\bar n\cdot q}\frac{\bar n\cdot p}{(\bar n\cdot p)(n\cdot q)+i\epsilon}\oneov{q^2-m_g^2+i\epsilon}    \,.\label{eq:36}
\end{align}
Equation (\ref{eq:36}) is scaleless and thus vanishes. The naive results corresponding to the diagrams of Figs.~\ref{fig:2}(b) and \ref{fig:2}(c) are
\begin{align}
i{\tilde m_b^n}&=(-im_0)(2g^2C_F)\delta_{\bar n\cdot p+\bar n\cdot \tilde q,Q}\delta_{\bar n\cdot p,Q}\mu^{2\epsilon}\nu^\eta
\int\!\frac{d^Dq}{(2\pi)^D}(-2\pi i)\delta(q^2)\frac{|\bar n\cdot q|^{-\eta}}{\bar n\cdot q}\frac{\bar n\cdot(p-q)}{(p\!-\!q)^2+i\epsilon}\delta(\bar n\cdot q_r-l^-) \label{eq:37}\\
i{\tilde m_c^n}&=(im_0)(2g^2C_F)\delta_{\bar n\cdot p+\bar n\cdot \tilde q,Q}\delta_{\bar n\cdot p,Q}\mu^{2\epsilon}(D-2)
\int\frac{d^Dq}{(2\pi)^D}(-2\pi i)\delta(q^2)\frac{(\bar n\cdot q)(n\cdot q)}{((p-q)^2+i\epsilon)^2}\delta(\bar n\cdot q_r-l^-) \,,\label{eq:38}
\end{align}
where $\tilde q$ is the large component of the collinear gluon momentum which obeys label
momentum conservation, and $q=\tilde q+q_r$ with $q_r$ being the soft residual momentum. In the $n$-collinear function, the $n$-collinear quarks only couple with $n$-collinear gluons, which means $n\cdot \tilde q=0$ and $n\cdot q=n\cdot q_r$.
The two Kronecker deltas in front of the integrals in both Eq.~(\ref{eq:37}) and Eq.~(\ref{eq:38}) force $\bar n\cdot\tilde q=0$, which implies that gluons emitted from initial to final state only have soft momentum. 
As a result, Eqs.~(\ref{eq:37}) and (\ref{eq:38}) can be reduced to
\begin{align}
i{\tilde m_b^n}&=(-im_0)(2g^2C_F)\delta_{\bar n\cdot \tilde q,0}\int\frac{d^Dq_r}{(2\pi)^D}(-2\pi i)\delta(q_r^2)
\frac{|\bar n\cdot q_r|^{-\eta}}{\bar n\cdot q_r}\frac{\bar n\cdot(p-q_r)}{(p-q_r)^2+i\epsilon}\delta(\bar n\cdot q_r-l^-)\label{eq:39}\\
i{\tilde m_c^n}&=(im_0)(2g^2C_F)\delta_{\bar n\cdot\tilde q,0}\int\frac{d^Dq_r}{(2\pi)^D}(-2\pi i)
\delta(q_r^2)\frac{(\bar n\cdot q_r)(n\cdot q_r)}{((p-q)^2+i\epsilon)^2}\delta(\bar n\cdot q_r-l^-)\,,\label{eq:40}
\end{align}
which is equal to the zero-bin subtraction. Therefore, after subtracting Eqs.~(\ref{eq:39}) and (\ref{eq:40}) from Eqs.~(\ref{eq:37}) and (\ref{eq:38}) respectively, the results vanish.

After computing the virtual collinear diagrams in Eqs.~(\ref{eq:35}) and (\ref{eq:36}) and adding their mirrors, we have 
to $\mathcal{O}(\alpha_s)$ 
\begin{align}
\sum m =\:&\:
{C}_n^{(0)}(Q-k) \, \frac{\alpha_s C_F}{\pi}w^2 
\bigg\{ \frac{e^{\epsilon \gamma_E} \Gamma(\epsilon)}{\eta}\bigg(\frac{\mu^2}{m^2_g}\bigg)^\epsilon +\frac{1}{\epsilon}\bigg[ 1 + \ln\frac{\nu}{\bn\cdot p} \bigg]
\nn\\&
+\ln\frac{\mu^2}{m_g^2}\ln\frac{\nu}{\bar n\cdot p}+\ln\frac{\mu^2}{m_g^2}+1-\frac{\pi^2}{6}\bigg\}\,,\label{diagsum}
\end{align}
which depends on the rapidity regulator. A natural choice of  $\nu\sim \bar n\cdot p=Q$ minimizes the rapidity logarithm. The collinear matrix element is obtained by multiplying the above result by the quark wave function renormalization
\begin{equation}
Z_{\xi} = 1 - \frac{\alpha_{s}C_{F}}{4 \pi} \bigg( \frac{1}{\epsilon} + \ln\frac{\mu^{2}}{m_g^2}+1\bigg)\,,
\end{equation}
which gives
\begin{align}
{C}_n^{(1)}(Q-k)=\:&\:
{C}_n^{(0)}(Q-k) \, \frac{\alpha_s C_F}{\pi}w^2 
\bigg\{ \frac{e^{\epsilon \gamma_E} \Gamma(\epsilon)}{\eta}\bigg(\frac{\mu^2}{m^2_g}\bigg)^\epsilon +\frac{1}{\epsilon}\bigg[ \frac{3}{4} + \ln\frac{\nu}{\bn\cdot p} \bigg]
\nn\\&
+\ln\frac{\mu^2}{m_g^2}\ln\frac{\nu}{\bar n\cdot p}+\frac{3}{4}\ln\frac{\mu^2}{m_g^2}+\frac{3}{4}-\frac{\pi^2}{6}\bigg\}\,.\label{eq:24temp}
\end{align}

\subsubsection{Soft function to $\mathcal{O}(\alpha_s)$ for DIS}
\begin{figure}
\centering
\includegraphics[scale=0.7]{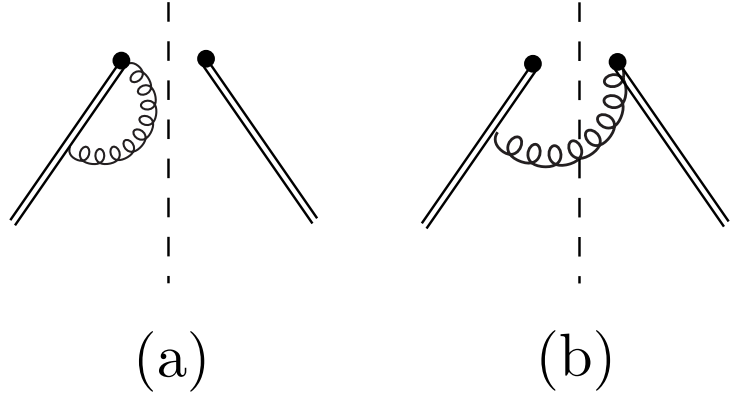}
\caption{$\mathcal{O}(\alpha_s)$ soft function Feynman diagrams: (a) is the virtual contribution; (b) is the real contribution.}
\label{fig:DISsoft}
\end{figure}

The soft function, given in \req{eq:07}, at tree level is
\begin{equation}\label{eq:13''}
S(l)^{(0)}=\delta(l).
\end{equation}
To $\mathcal{O}(\alpha_s)$, with the $\eta$-regulated soft Wilson line Eq.\,\eqref{Wsofteta}, we can explicitly isolate the rapidity poles of the soft function. The Feynman diagrams for the one-loop soft functions are shown in Fig.\,\ref{fig:DISsoft}, where Fig.\,\ref{fig:DISsoft}(a) is the virtual piece and Fig.\,\ref{fig:DISsoft}(b) is the real piece.
The double lines represent the eikonal lines. Here we also omit the mirror images of Figs.\,\ref{fig:DISsoft}(a) and 3(b).

The naive virtual soft function amplitude determined from Fig.\,\ref{fig:DISsoft}(a) is 
\begin{align}
\tilde S_v&=(2ig^2C_F)\delta(l)\mu^{2\epsilon}\nu^\eta w^2 
\int d^dk\frac{|2k_3|^{-\eta}}{k^2-m_g^2+i\epsilon}\oneov{k^-+i\epsilon}\oneov{k^++i\epsilon}\nn \\
&=\delta(l)\frac{\alpha_sC_F}{\pi}w^2\bigl[-\frac{e^{\epsilon\gamma_E}\Gamma(\epsilon)}{\eta}\paren{\frac{\mu}{m_g}}^{2\epsilon}+\oneov{2\epsilon^2}
+\oneov{\epsilon}\ln\frac{\mu}{\nu}+\ln^2\frac{\mu}{m_g}-\ln\frac{\mu^2}{m_g^2}\ln\frac{\nu}{m_g}-\frac{\pi^2}{24}\bigl]. \label{eq:44}
\end{align}
The zero-bin subtraction for the naive virtual piece is the overlap with the $n$ and $\bn$-ollinear directions,
\begin{align}
S_{v\phi}^{\bar n}&(k^-\gg k^+)=(2ig^2C_F)\delta(l)\mu^{2\epsilon}\nu^\eta\int\frac{d^Dk}{(2\pi)^D}
\frac{|k^-|^{-\eta}}{(k^++i\epsilon)(k^-+i\epsilon)(k^2-m_g^2+i\epsilon)}\,,\label{eq:45}\\
S_{v\phi}^{n}&(k^+\gg k^-)=(2ig^2C_F)\delta(l)\mu^{2\epsilon}\nu^\eta\int\frac{d^Dk}{(2\pi)^D}
\frac{|k^+|^{-\eta}}{(k^++i\epsilon)(k^-+i\epsilon)(k^2-m_g^2+i\epsilon)}.\label{eq:46}
\end{align} 
These integrals are scaleless in rapidity regularization and vanish. This must be the case because adding the rapidity regulator to the soft Wilson lines Eq.\eqref{Wsofteta} restricts the soft function integral to lie only in the soft momentum region.  In other words, in the virtual contributions, the rapidity regulator properly separates soft and collinear modes in $\text{SCET}_{\text{II}}$.   Thus the total virtual soft function is
\begin{equation}
S_v=2\tilde S_v\,.
\end{equation}

The naive real contribution from the diagram in Fig.\,\ref{fig:DISsoft}(b) is
\begin{align}
\label{realint}
\tilde S_\textrm{r} &= + 4\pi C_F g^2_s \mu^{2 \epsilon} w^2 \nu^\eta \int \frac{d^Dk}{(2\pi)^{D-1}}\delta(k^2-m^2_g) \delta(\ell-k^+)|2 k_3|^{-\eta} \frac{1}{k^+}\frac{1}{k^-} 
\\&= -\frac{\alpha_s C_F}{\pi}\bigg(e^{ \gamma_E} \frac{\mu^2}{m^2_g}\bigg)^\epsilon w^2 \nu^\eta \frac{\theta(\ell)}{\ell^{1+\eta}} \Gamma(\epsilon)\,.\nn
\end{align}
In the scheme introduced in our previous paper~\cite{Fleming:2012kb} the collinear zero-bin subtraction for the real soft function is given by expanding the real soft contribution about the collinear limit everywhere in the integrand except in the measurement function. Then our zero-bin subtraction is not 0 at this order, because overlap with the collinear regions in the soft function is not suppressed by the rapidity regulator in the initial state Wilson lines.  Mathematically, we see this by the presence of the scale brought into the integral by the measurement function.  The overlap of the integral in Eq.~(\ref{realint}) with the $n$-collinear region is given by taking the limit $k^+\gg k^-$ with $k^+k^- \sim k^2_\perp$,
\begin{align}
\label{sn}
S_{n\phi} ^{r}&=  - 4\pi C_F g^2_s \mu^{2 \epsilon} w^2 \nu^\eta \int \frac{d^Dk}{(2\pi)^{D-1}}\delta(k^2-m^2_g) \delta(\ell-k^+)|k^+|^{-\eta} \frac{1}{k^+}\frac{1}{k^-}  \\
& = +\frac{\alpha_s C_F}{\pi}\bigg(e^{ \gamma_E} \frac{\mu^2}{m^2_g}\bigg)^\epsilon w^2 \nu^\eta \frac{\theta(\ell)}{\ell^{1+\eta}} \Gamma(\epsilon)\,,\nn
\end{align}
which is the same as the result in Eq.~(\ref{realint}). The $\bar n$-collinear subtraction is given by taking the limit $k^-\gg k^+$ with $k^+k^- \sim k^2_\perp$ in the first line of Eq.~(\ref{realint}),
\begin{align}
\label{sbarn}
S_{\bar n\phi} ^{r}&=  - 4\pi C_F g^2_s \mu^{2 \epsilon} w^2 \nu^\eta \int \frac{d^Dk}{(2\pi)^{D-1}}\delta(k^2-m^2_g) \delta(\ell-k^+)|k^-|^{-\eta} \frac{1}{k^+}\frac{1}{k^-}  \\
&= -\frac{\alpha_s C_F}{\pi}\bigg(e^{ \gamma_E} \frac{\mu^2}{m^2_g}\bigg)^\epsilon w^2 \bigg(\frac{\nu}{m^2_g}\bigg)^\eta \frac{\theta(\ell)}{\ell^{1-\eta}} 
\frac{\Gamma(\eta+\epsilon)}{\Gamma(1+\eta)}\,.\nn
\end{align}
Comparing Eqs.~\req{realint}-\req{sbarn}, we see that the unsubtracted soft function $\tilde S_r$ is dominated by overlap with the $n$-collinear region as \req{sn} represents the $n$-collinear modes running into the soft function.  This is due to the measurement being on soft radiation only in the $n$-collinear direction.  Radiation in the $\bar n$-collinear direction has been integrated out in the matching onto $\text{SCET}_{\text{II}}$ and subtracting \req{sbarn} from \req{realint} removes the momentum in the soft function that overlaps with the $\bar n$-collinear momentum region.  Thus the zero-bin subtracted real contribution, given by the diagrams in Fig.\,\ref{fig:DISsoft}(b), is
\begin{align}
\label{softreal}
S_\textrm{r} &= 2(\tilde S_\textrm{r}-S_{n\phi} ^{r} -S_{\bar n\phi} ^{r})= - 2S_{\bar n\phi} ^{r} \\
&= 2 \frac{\alpha_s C_F}{\pi} w^2\frac{1}{Q} \bigg\{\bigg[\frac{1}{2} \frac{e^{\epsilon \gamma_E} \Gamma(\epsilon)}{\eta}\bigg(\frac{\mu}{m_g}\bigg)^{2\epsilon}
-\frac{1}{2\epsilon^2} +\frac{1}{2 \epsilon} \ln\frac{\nu Q}{\mu^2}-\ln^2\frac{\mu}{m_g} + \ln\frac{\mu}{m_g}\ln\frac{\nu Q}{m^2_g}+\frac{\pi^2}{24}\bigg]\delta(z)\nn\\
& \hspace{8ex}+\bigg[\frac{1}{2\epsilon}+\ln\frac{\mu}{m_g}\bigg] \left(\frac{1}{z}\right)_+\bigg\} \,, \nn
\end{align}
where the plus function of the dimensionful variable $\ell$ is given in terms of the definition of a dimensionless variable $z = \ell/\kappa$,
\begin{align}
\left(\frac{1}{\ell}\right)_+ = \frac{1}{\kappa}\left(\frac{1}{z}\right)_+ + \ln \kappa \, \delta(\kappa \,z) \,,
\end{align}
with
\begin{align}
\left(\frac{1}{z}\right)_+ \equiv \lim_{\beta \to 0} \bigg[ \frac{\theta(z-\beta)}{z}+\ln \beta \, \delta(z)\bigg] \,. 
\end{align}

Adding the virtual and real contributions gives the one-loop expression for the soft function
\begin{eqnarray}
\label{finalsoft}
S(z)^{(1)} &=& \frac{\alpha_s C_F}{\pi}w^2\frac{1}{Q}\bigg\{ -\frac{e^{\epsilon \gamma_E} \Gamma(\epsilon)}{\eta}\bigg(\frac{\mu}{m_g}\bigg)^{2\epsilon} \delta(z)+ \bigg( \frac{1}{\epsilon}+ \ln\frac{\mu^2}{m^2_g}\bigg)
\bigg(\!-\ln\frac{\nu}{Q}\delta(z)+\left(\frac{1}{z}\right)_+\bigg) \bigg\}. \label{eq:13}
\end{eqnarray}
Logarithms in the soft function are minimized by setting $\mu\sim m_g$ and $\nu\sim\ell\sim Qz\sim Q\paren{\frac{1-x}{x}}$.  Note that $Q \paren{\frac{1-x}{x}}$ is an end-point region energy scale, which is however different from what one naturally chooses for the collinear function. Clearly, resumming logarithms in $\nu$ is needed. 

At this point we wish to alert the reader to an alternative approach to deriving Eq.~(\ref{finalsoft}), developed in Ref.~\cite{Hoang:2015iva}. In our work we strictly take the $m_{g}\to 0$ limit while holding the momentum $\ell$ fixed in the soft contribution, and determine the zero-bin subtraction as outlined above. In contradistinction, the authors of Ref.~\cite{Hoang:2015iva} hold $m_{g}$ fixed and consider both  $\ell>m_g$ and $\ell<m_g$, and then take the $m_{g} \to 0$ limit in the soft contribution. The zero-bin subtractions are determined by expanding the soft integrand around the collinear limit, including the measurement function but excluding the rapidity regulator term $|2 \ell_{3}|^{-\eta}$.  These two approaches result in different collinear zero-bin subtractions for the soft function; while we have both  an $n$-collinear and $\bn$-collinear subtraction, the approach of Ref.~\cite{Hoang:2015iva} requires no collinear zero-bin subtraction in the soft function. In DIS at the end point the two approaches give the same results up to $\mathcal{O}(m_g/\ell)$, which vanishes in the $m_{g}\to 0$ limit. Thus, there is no way to determine from DIS if one of the two (or both) of the approaches is inconsistent. However, as we point out in Sec.~\ref{sec:sec3}, DY cannot be treated with our approach, while the approach used in Ref.~\cite{Hoang:2015iva} gives a consistent result. Furthermore, our approach is not compatible with the threshold expansion while that in Ref.~\cite{Hoang:2015iva} is~\cite{HoangPrivate}.

\subsubsection{Renormalization group running for DIS}

To subtract the divergences in $\epsilon$ and $\eta$ in Eqs.~(\ref{eq:24temp}) and (\ref{finalsoft}), we introduce counterterms
\begin{align}
{C}_{n}( Q-k)^{R} &=Z_{n}^{-1} {C}_{n}( Q-k)^{B}\,, \nn\\
S(\ell)^{R} &= \int dz' Z_s(z-z')^{-1}S(\ell')^{B} \,, \nn
\end{align}
where $\ell'=Qz'$ and superscripts $R$ and $B$ indicate renormalized and bare. The one-loop collinear counterterm is
\beq
\label{collct}
Z_{n} =1 + \frac{ \alpha_s C_F}{\pi}w^2\bigg[\frac{e^{\epsilon \gamma_E} \Gamma(\epsilon) }{ \eta}\bigg(\frac{\mu}{m_g}\bigg)^{2 \epsilon}+
\frac{1}{\epsilon}\bigg( \frac{3}{4} + \ln \frac{\nu}{\bn\cdot p}\bigg)\bigg]\,,
\eeq
and the one-loop soft counterterm is 
\beq
\label{softct}
Z_s(z) = 
\delta(z) + \frac{\alpha_s C_F}{\pi}w^2\bigg\{- \frac{e^{\epsilon \gamma_E} \Gamma(\epsilon) }{ \eta}\bigg(\frac{\mu}{m_g}\bigg)^{2 \epsilon}\delta(z)
+\frac{1}{\epsilon} \bigg[\left(\frac{1}{z}\right)_+ -\ln\frac{\nu}{Q}\delta(z)\bigg]\bigg\}\,.
\eeq
These counterterms obey the consistency condition put forth in Ref.\,\cite{Fleming:2007xt}, as they must,
\beq\label{consistencycond}
Z_H Z_{J_{\bn}}(z) = Z_{n}^{-1} Z_s^{-1}(z)\,,
\eeq
where $Z_{J_{\bn}}(z)$ is the jet-function counterterm and $Z_H$ is the square of the counterterm for the SCET DIS current, which has been given at one loop in Ref.~\cite{bauer03} in $4-\epsilon$ dimensions.   Converting the result of  Ref.~\cite{bauer03} to $4 - 2 \epsilon$ dimensions and squaring gives
\beq
Z_H = 1 - \frac{\alpha_s C_F}{2\pi} \bigg( \frac{2}{\epsilon^2}+ \frac{3}{\epsilon}+ \frac{2}{\epsilon}\ln \frac{\mu^2}{Q^2}\bigg)\,,
\label{ZH}
\eeq
where $Q^2 = \bn \cdot p\, n\cdot p_X$.  
The one-loop result for $Z_{J_{\bn}}(z)$ is given by Ref.~\cite{Manohar:2003vb},
\beq
Z_{J_{\bn}}(z) = \delta(z) +  \frac{\alpha_s C_F}{4\pi}\bigg[ \bigg(\frac{4}{\epsilon^2}+\frac{3}{\epsilon}-\frac{1}{\epsilon}\ln\frac{(n\cdot p)Q}{\mu^2}\bigg)\delta(z) -\frac{4}{\epsilon}\left(\frac{1}{z}\right)_+\bigg] \,.
\eeq
Putting the factors together,
\beq\label{ctconsistent}
Z_H Z_{J_{\bn}}(z)=   
\delta(z) +  \frac{\alpha_s C_F}{4\pi}\bigg\{ \bigg[-\frac{3}{\epsilon}+\frac{4}{\epsilon}\ln \left(\frac{\bn\cdot p}{Q}\right)\bigg]\delta(z) -\frac{4}{\epsilon}\left(\frac{1}{z}\right)_+\bigg\} \,,
\eeq
which is exactly equal to the product of inverses $Z_{n}^{-1} Z_s^{-1}(z)$ taken from Eqs.~(\ref{collct}) and (\ref{softct}).

From the one-loop results, we extract the $\mu$ anomalous dimensions for the collinear and soft function respectively,
\begin{eqnarray}\label{anomdimmuDIS}
\gamma^\mu_{n} (\mu,\nu)&=& \frac{2 \alpha_s(\mu) C_F}{\pi}\bigg(\frac{3}{4}+\ln\frac{\nu}{\bn\cdot  p}\bigg)\\
\gamma^\mu_s(\mu,\nu) &=& \frac{2 \alpha_s(\mu) C_F}{ \pi}\bigg[\left(\frac{1}{z}\right)_+ -\ln \frac{\nu}{Q} \delta(z)\bigg] \,.\nn
\end{eqnarray}
Note that
\begin{equation}\label{eq:15'}
\gamma^\mu=\gamma_n^\mu \delta(z)+\gamma_s^\mu=\frac{2\alpha_s C_F}{\pi}\left\{\left[\frac34-\ln\left(\frac{\bar n\cdot p}{Q}\right)\right]\delta(z)+\left(\frac{1}{z}\right)_+\right\}   \,,
\end{equation}
which agrees with the known result, and the $\nu$-dependence cancels as expected. In Mellin moment space this is the $n=0$ result given in Eq.~(\ref{mellmomanom}). We can now trace the origin of the large logarithm to the rapidity region. If we choose $\nu = \nu_{c}\sim Q$ in the collinear anomalous dimension on the first line of Eq.~(\ref{anomdimmuDIS}) and $\nu = \nu_{s}\sim Q(1-x)$ in the soft anomalous dimension in the second line, then neither term contains large logarithms. Adding the two anomalous dimensions together then gives
 \begin{equation}\label{newanomdim}
\gamma^\mu=\gamma_n^\mu\delta(z)+\gamma_s^\mu=\frac{2\alpha_s C_F}{\pi}\left\{\left[\frac{3}{4}-\ln\left(\frac{\nu_{s}}{\nu_{c}}\frac{\bar n\cdot p}{Q}\right)\right]\delta(z)+\left(\frac{1}{z}\right)_+\right\}   \,,
\end{equation}
where the combination of plus-function and logarithmic term is no longer anomalously enhanced compared to the 3/4.

Minimizing the logarithmic term in the $\mu$ anomalous dimension requires choosing two widely separated rapidity scales, $\nu_{c}$ and $\nu_{s}$. This necessitates a resummation of logarithms of $\nu$. The $\nu$ anomalous dimensions for the collinear and soft functions are
\begin{align}
\gamma_n^\nu(\mu,\nu)&=\frac{\alpha_s(\mu)C_F}{\pi}\ln\frac{\mu^2}{m_g^2}   \,,\nn\\
\gamma_s^\nu(\mu,\nu)&=-\frac{\alpha_s(\mu)C_F}{\pi}\ln\frac{\mu^2}{m_g^2}\delta(z)   \,.\label{anomdimnuDIS}
\end{align}
Adding them together, we have
$\gamma^\nu=\gamma_n^\nu \delta(z)+\gamma_s^\nu=0$, as is dictated by the consistency condition.  The presence of $m_g$ in $\gamma_n^\nu$ and $\gamma_s^\nu$ indicates that the renormalization group running in $\nu$ depends on an infrared scale, and therefore is nonperturbative. Thus we are left with little choice but to treat the $\nu$ resummation as 
part of the nonperturbative aspect of DIS and to absorb it into the definition of the PDF.

The  $\mu$ and $\nu$ running are independent and can be carried out in either order; however they must obey the constraint
\begin{equation}
\label{constraint}
\frac{d}{d \ln \mu} \gamma^{\nu} = \frac{d}{d\ln \nu}\gamma^{\mu}\,.
\end{equation} 
For the collinear function, the $\mu$ running is given to one loop by
\begin{align}\label{Cnmurunning}
{C}_{n}(Q-k;\mu,\nu_c) &= U(\mu,\mu_0,\nu_c){C}_{n}(Q-k;\mu_0,\nu_c) \\
U(\mu,\mu_0,\nu_c)&=e^{\frac{3}{4}\omega(\mu_{0},\mu)} \bigg[\frac{\nu_c}{\bn\cdot p}\bigg]^{\omega(\mu_{0},\mu)}
\,, \nn
\end{align}
where $\nu_c$ is the collinear rapidity scale and
\beq
\omega(\mu_{0},\mu) = \frac{4 C_F}{\beta_0}\ln\bigg[\frac{\alpha_s(\mu_{0})}{\alpha_s(\mu)}\bigg]\,.
\eeq
Note that $\omega(\mu_{0},\mu) = 2 a_{\Gamma}(\mu,\mu_{0})$ of Ref.~\cite{Becher:2006mr}.
For the soft function, the one-loop $\mu$ running is
\begin{align}\label{Smurunning}
S(\ell;\mu,\nu_s) &= \int d r \, U(\ell-r;\mu,\mu_0,\nu_s) S(\ell;\mu_0,\nu_s)\\
 U(\ell-r;\mu,\mu_0,\nu_s)&=\frac{ \big( e^{ \gamma_E}\nu_s\big)^{-\omega(\mu_{0},\mu)}}{\Gamma(\omega(\mu_{0},\mu))}\left(\frac{1}{(\ell-r)^{1- \omega(\mu_{0},\mu)}}\right)_+\,. \nn
\end{align}
Combining the running factors we find
\begin{equation}
\label{combinedrun}
U(\mu,\mu_0,\nu_c)U(\ell-r;\mu,\mu_0,\nu_s) =  \bigg[ \frac{e^{- \gamma_E}\nu_c}{\bn\cdot p \, \nu_{s}}\bigg]^{\omega(\mu_{0},\mu)}
\frac{e^{\frac{3}{4}\omega(\mu_{0},\mu)} }{\Gamma(\omega(\mu_{0},\mu))}\left(\frac{1}{(\ell-r)^{1- \omega(\mu_{0},\mu)}}\right)_+\,.
\end{equation}
This agrees with Eq.~(66) of Ref.~\cite{Becher:2006mr} if we set $\nu_{c} = \nu_{s}$, convert the plus-distribution to dimensionless variables, and recognize that $2 a_{\gamma^{\phi}}(\mu_{f},\mu_{0})= (3/4) \omega(\mu_{0},\mu_{f})$ at this order.

To get a  feel for which logarithms are being summed we will transform the combined running factors into Mellin moment space (for large $N$),
\begin{equation}
\label{mellinrunfac}
U(\mu,\mu_0,\nu_c)U(N;\mu,\mu_0,\nu_s) =  \bigg[ \frac{e^{- \gamma_E}\nu_c}{\bar{N} \, \nu_{s}}\bigg]^{\omega(\mu_{0},\mu)}e^{\frac{3}{4}\omega(\mu_{0},\mu)} \,.
\end{equation}
The first term on the right-hand  side in square brackets can be expressed as
\begin{eqnarray}
\label{logseries}
 \bigg[ \frac{e^{- \gamma_E}\nu_c}{\bar{N} \, \nu_{s}}\bigg]^{\omega(\mu_{0},\mu)} &=& \textrm{Exp}\bigg[ \omega(\mu_{0},\mu) \ln\bigg( \frac{e^{- \gamma_E}\nu_c}{\bar{N} \, \nu_{s}}\bigg)\bigg]\nn\\
&=& \textrm{Exp}\bigg[ \frac{4 C_{F}}{\beta_{0}}\ln\bigg( \frac{\bar{N} \, \nu_{s}}{e^{- \gamma_E}\nu_c}\bigg) \sum^{\infty}_{n=1} \frac{1}{n}\bigg(\frac{\beta_{0} \alpha_{s}(\mu)}{2 \pi} \ln\frac{\mu}{\mu_{0}}\bigg)^{n}\bigg],
\end{eqnarray}
which, in the exponent,  gives a series in $\alpha^{n}_{s}(\mu)\ln^{n}(\mu/\mu_{0})$ times a single power of $\ln(\bar{N}\nu_{s}/\nu_{c})$. If we make the choice $\nu_{c}=\nu_{s}$ we reproduce
the standard result of a single logarithmic series multiplied by a single logarithm of $N$. However, if we make the choice for $\nu_{c}$ and $\nu_{s}$ given above then we merely have a single logarithmic series multiplied by an ${\mathcal O}(1)$ quantity. We argue that this is the natural choice from an EFT perspective.

Having widely separated rapidity scales then forces us to consider the rRGE.  Although the $\nu$ running is nonperturbative it is still enlightening to see what the resummation looks like, and we push ahead and determine the soft $\nu$ running factor using the constraint Eq.~(\ref{constraint}) to sum large logarithms in the rapidity anomalous dimension,
\begin{align}
\label{softnurun}
S(\ell;\mu_s,\nu)&= V(\mu_s, \nu,\nu_0) S(\ell;\mu_s,\nu_0),   \\
V(\mu_s, \nu,\nu_0)&= \bigg[\frac{\nu}{\nu_0}\bigg]^{\omega(\mu_{s},m_{g})} \,.\nn
\end{align}
Note that if we choose $\nu = \nu_{c}$ and $\nu_{0}=\nu_{s}$ in the above equations with $\mu_{s}=\mu$ then 
\begin{equation}
V(\mu, \nu_{s},\nu_c) = \bigg[\frac{\nu_{c}}{\nu_s}\bigg]^{\omega(\mu,m_{g})} = \bigg[\frac{\nu_{c}}{\nu_s}\bigg]^{\omega(\mu,\mu_{0})}\bigg[\frac{\nu_{c}}{\nu_s}\bigg]^{\omega(\mu_{0},m_{g})} = \bigg[\frac{\nu_{s}}{\nu_c}\bigg]^{\omega(\mu_{0},\mu)} \bigg[\frac{\nu_{c}}{\nu_s}\bigg]^{\omega(\mu_{0},m_{g})}\,.
\end{equation}
The first term in square brackets on the far right-hand side cancels the $\nu_{s}/\nu_{c}$ dependence in Eq.~(\ref{combinedrun}), and results in a running factor identical to the one obtained without rapidity resummation. However, the second term in square brackets on the far right-hand side of this equation remains. This term is infrared sensitive and is absorbed into the definition of the PDF. 
Finally we expressed the leftover rapidity running factor as
\begin{equation}
\label{nonpertrrge}
V(\mu_{0}, \nu_{s},\nu_c)= \textrm{Exp}\bigg[ -\frac{4 C_{F}}{\beta_{0}}\ln\bigg( \frac{\nu_{c}}{\nu_s}\bigg) \sum^{\infty}_{n=1} \frac{1}{n}\bigg(\frac{\beta_{0} \alpha_{s}(\mu_{0})}{2 \pi} \ln\frac{\mu_{0}}{m_{g}}\bigg)^{n}\bigg]\,,
\end{equation}
with $\nu_{c}/\nu_{s} = \bar{N}$, making it clear that what is being summed (in Mellin moment space) by the rRGE is the product $\alpha^{n}_{s}(\mu_{0}) \ln^{n}(\mu_{0}/m_{g})\ln N$. The large logarithm of $N$ multiplies infrared logarithms, which explains why no one has tried to sum these terms before. 

Of course, this begs the question of why we should even bother to separate collinear from soft in the PDF. One answer is that we have a consistent EFT formalism that never produces terms that violate power counting. There are, however, more. Currently fits of the PDF produce a very steeply falling function of momentum fraction as the end point is approached, with no understanding of why; our result offers an explanation. To see why we define our PDF for large $x$ in DIS as a modified form of the function $f^{ns}_{q}$ in Eq.~(\ref{eq:011}),
\begin{equation}
f_q^{ns}(z;\mu)_{\textrm{end point}}= \delta_{\tilde n\cdot\tilde p,Q}\mathcal{Z}_n(\mu,\nu_{c})S^{(DIS)}(\ell;\mu,\nu_{s})V(\mu_{0}, \nu_{s},\nu_c)
\,. \label{eq:016}
\end{equation}
This is the same as the operator definition we give in our previous paper, but we have made the presence of the $V(\mu_{0}, \nu_{s},\nu_c)$ factor explicit. Away from the end point $\nu_{c}$ and $\nu_{s}$ must flow together so the PDF in the end point matches smoothly onto the usual definition of the PDF.  Choosing to set the rapidity scales in Mellin space with  $\nu_{c}/\nu_{s} = \bar{N}$, we have 
\begin{equation}
V(\mu_{0}, \nu_{s},\nu_c) = \bar{N}^{-\omega(m_{g},\mu_{0})}\,.
\end{equation}
If we transform back into momentum fraction space we find
\begin{equation}
V(\mu_{0}, \nu_{s},\nu_c)  =\frac{1}{\Gamma\big(\omega(m_{g},\mu_{0}) \big)} (1-z)^{\omega(m_{g},\mu_{0})-1}\,,\nn
\end{equation}
where the exponent of $(1-z)$ is nonperturbative and could be large. Thus we can interpret the conventional running of the PDF in the end point using the anomalous dimension in Eq.~(\ref{mellmomanom}) as a combined running in $\mu$ and in $\nu$, with a subset of potentially large nonperturbative rapidity logarithms remaining in the PDF. These remaining logarithms could then be responsible for the steep falloff of the PDF in the end point.

Finally, it is interesting to see how the above modification to the PDF fairs in the analysis carried out in Sec.~3.5 of Ref.~\cite{Becher:2006mr}. Nothing in that analysis changes if we identify 
\begin{equation}
b(\mu_{0}) = b_{\textrm{\tiny IR}} + \omega(m_{g},\mu_{0})\,,
\end{equation}
with $b_{\textrm{\tiny IR}}$ being the nonperturbative value of the $b$-parameter with absolutely no running. Furthermore, 
the relation for ${\mathcal N}(\mu)$ remains unchanged.

\subsubsection{Comparing to the perturbative QCD result}
In this section, we compare the one-loop expression of the hadronic tenor in SCET to that in QCD.  This provides a powerful check that nothing has been missed in the SCET calculation.
Extracting the scalar part of the SCET effective hadronic tensor from Eq.\eqref{hadtenSCET2}, we have
\begin{equation}
W_{\text{eff}}^{\mu\nu}=-\frac{g_\perp^{\mu\nu}}{2}W_{\text{eff}}
\end{equation}
where
\begin{equation}\label{eq:50''''}
W_{\text{eff}}=2QH(Q;\mu_q,\mu_c)\int_x^1 \frac{dw}{w}J_{\bar n}(Qw;\mu_c;\mu)C_n((Q-k);\mu_c;\mu,\nu)S^{\text{DIS}}(Q(1-w);\mu,\nu)\,.
\end{equation}
The renormalized hard function $H^R(Q;\mu_q;\mu_c)$ and jet function $J^R_{\bar n}(Qz;\mu_c;\mu)$ are given in the literature \,\cite{Manohar:2003vb, Chay:2013zya, bauer03, Becher:2006mr, Hornig2009, Bauer2010},
\begin{align}
H_{\text{DIS}}^R(Q,\mu)&=1+\frac{\alpha_sC_F}{2\pi}\left(-\ln^2\frac{\mu^2}{Q^2}-3\ln\frac{\mu^2}{Q^2}-8+\frac{\pi^2}{6}
\right)\label{eq:50'}\\
J_{\bar n}^R(Q(1-x),\mu)&=\delta(1-x)+\frac{\alpha_sC_F}{2\pi}\bigg\{
\delta(1-x)\left(\frac32 \ln\frac{\mu^2}{Q^2}+\ln^2\frac{\mu^2}{Q^2}+\frac72-\frac{\pi^2}{2}\right)\nn\\
&-\left(\frac{2}{1-x}\right)_+\left(\ln\frac{\mu^2}{Q^2}+\frac34\right)+2\left(\frac{\ln(1-x)}{1-x}\right)_+
\bigg\}   \,.\label{eq:50''}
\end{align}
From Eqs.~\req{eq:24temp} and \req{collct}, we obtain the renormalized collinear function,
\begin{align}
C^R(Q-k;\mu,\nu)&=m_0\delta_{\bar n,\tilde p,Q}\delta(k)\bigg[1+\frac{\alpha_sC_F}{\pi}\bigg(\ln\frac{\mu^2}{m_g^2}\ln\frac{\nu_{c}}{\bar n\cdot p} 
+\frac34\ln\frac{\mu^2}{m_g^2}+\frac34-\frac{\pi^2}{6}
\bigg)
\bigg]  \,.\label{eq:50'''}
\end{align}
From Eqs.~\req{finalsoft} and \req{softct}, we obtain the renormalized soft function,
\begin{equation}\label{eq:50a}
S^R(Q(1-x);\mu,\nu)=\frac1Q\delta(1-x)+\frac{\alpha_sC_F}{\pi Q}\left\{\ln\frac{\mu^2}{m_g^2}\left[
\left(\frac{1}{1-x}\right)_+-\ln\frac{\nu_{s}}{Q}\delta(1-x)
\right]
\right\}.
\end{equation}
Inserting Eqs.\,\eqref{eq:50'}-\eqref{eq:50a} into \eqref{eq:50''''}, we arrive at the one-loop expression for the hadronic structure function calculated in SCET which is valid in the end-point region:
\begin{align}
W_{\text{eff}}&=2m_0\delta_{\bar n,\tilde p,Q}\bigg\{\delta(1-x)
+\frac{\alpha_sC_F}{\pi}\bigg[\left(
-\frac34\ln\frac{m_g^2}{Q^2}-\frac32-\frac{\pi^2}{3}
\right)\delta(1-x)\nn\\
&-\left(\frac{1}{1-x}\right)_+\left(\ln\frac{m_g^2}{Q^2}+\frac34\right)+\left(\frac{\ln(1-x)}{1-x}\right)_+
+\ln \frac{\mu^{2}}{m_{g}^{2}} \ln\frac{\nu_{c}}{\nu_{s}}  \bigg]
\bigg\}\,.
\label{eq:50b}
\end{align}
Note the rapidity scale dependence in the last term is multiplied by an IR logarithm indicating once again that the logarithms that are being summed are infrared in nature. In order to compare to $W$ in QCD we first set $\nu_{c}=\nu_{s}$.

The quark contribution to the hadronic structure function in  perturbative QCD is given in Ref.\,\cite{Field1989},
\begin{align}
\mathcal{F}_2(x)&=\int_x^1\frac{dy}{y}(G_{p-q}^{(0)}(y)+G_{p-\bar q}^{(0)}(y))\bigg\{
\delta(1-z) 
+\frac{\alpha_s}{2\pi}P_{q\to gq}(z)\ln\frac{Q^2}{m_g^2}+\alpha_sf_2^{q \text{ DIS}}(z)\bigg\}\,,\label{eq:50c}
\end{align}
where
\begin{align}
P_{q\to qg}(z)&=\frac43\left(\frac{1+z^2}{1-z}\right)_+  =\frac{4}{3}\left((1+z^2)\left(\frac{1}{1-z}\right)_+ +\frac{3}{2}\delta(1-z)\right),\nn\\
\alpha_sf_2^{q\text{ DIS}}(z)&=\frac{2\alpha_s}{3\pi}\bigg[(1+z^2)\left(\frac{\ln(1-z)}{1-z}\right)_+-\frac{1+z^2}{1-z}(2\ln z)-\frac32\left(\frac{1}{1-z}\right)_+\nn\\
&+4z+1-\left(\frac{2\pi^2}{3}+\frac94\right)\delta(1-z)\bigg],\label{eq:50d}
\end{align}
$z=x/y$, and $G_{p\to q}^{(0)}+G_{p\to \bar q}^{(0)}=2m_0\delta_{\bar n,\tilde p,Q}$.
As $x\to 1$, we have
\begin{align}
\mathcal{F}_2\xrightarrow{z\to 1}&\left(2m_0\delta_{\bar n\cdot \bar p,Q}\right)\bigg\{ \delta(1-x)+\frac{\alpha_sC_F}{\pi}\bigg[
\left(-\frac34\ln\frac{m_g^2}{Q^2}-\frac38 - \frac{\pi^2}{3}\right)\delta(1-x)\nn\\
&
-\left(\frac{1}{1-x}\right)_+\bigg(\ln\frac{m_g^2}{Q^2}+\frac34\bigg)+\left(\frac{\ln(1-x)}{1-x}\right)_++9\bigg]\bigg\}.\label{eq:50e}
\end{align}
Comparing Eq.~\req{eq:50e} to Eq.~\req{eq:50b}, we find that the low energy behavior agrees. In particular, by comparing the
jet function and soft function separately in SCET, we can trace the origin of the $m_g^2$ dependence in the quark splitting term $\sim P_{q\to qg}\ln\frac{Q^2}{m_g^2}$ to the large scale difference between the collinear gluons and the soft gluons entering the final state jet. The difference between Eq.\,\eqref{eq:50e} and Eq.\,\eqref{eq:50b} is the constant coefficient of $\delta(1-x)$ and the constant term. The former is regularization scheme dependent, and the latter subleading.  Since the SCET calculation uses a different regularization scheme from Ref.\,\cite{Field1989} this discrepancy is expected.

\section{Drell-Yan at end point with Rapidity Regulator}\label{sec:sec3}

We now apply a similar analysis to the Drell-Yan processes. We investigate DY in the semi-inclusive region of phase space where the momentum fractions $x,\bar x$ of the two colliding partons become large, approaching the maximal value $x\sim\bar x\sim 1$.  
Drell-Yan in the large-$x$ region has been investigated before using perturbative QCD factorization techniques \cite{Akhoury1998,catani1989,Sterman1986,korchemsky87,korchemsky93} as well as effective field theory techniques based on SCET~\cite{becher07}.
Although the end point in Drell-Yan is not accessible in real experiments, it is of theoretical interest to investigate how the parton distribution functions in two protons interfere with each other at large $x$. 
 
We analyze Drell-Yan at threshold by integrating out the large scale $\sim Q$ by matching QCD onto $\text{SCET}_{\text{II}}$, and then we factorize.  We compute each piece in the factorization formula to the first perturbative order and resum large logarithms to NLL order.  Finally, we discuss the PDF for two protons colliding at large $x$.

\subsection{Kinematics}\label{sec:sec3.1}
While we worked through the kinematics of the DIS process in both the target rest frame and Breit frame, we consider the Drell-Yan process only in the Breit frame. The proton in the $\bar n$ direction  carries momentum $p^\mu=\frac{\bar n^\mu}{2}n\cdot p+\frac{n^\mu}{2}\bar n\cdot p+\bar p_\perp^\mu$, and the proton in the $n$ direction  carries momentum $\bar p^\mu=\frac{n^\mu}{2}\bar n\cdot \bar p+\frac{\bar n^\mu}{2}n\cdot \bar p+\bar p_\perp^\mu$. The invariant mass squared of the proton-proton collision is $s=(p+\bar p)^2\simeq (n\cdot p)(\bar n\cdot \bar p)$, since $n\cdot p$ and $\bar n\cdot \bar p$ are the large components of $p$ and $\bar p$ respectively. The squared momentum transfer between the two protons is $Q^2=q^2$, so for $p$ we define
\begin{equation}\label{eq:16'}
x=\frac{Q^2}{2p\cdot q}=\frac{Q^2}{(n\cdot p)(\bar n\cdot q)}\simeq \frac{n\cdot q}{n\cdot p}\,,
\end{equation}
while for $\bar p$  we define
\begin{equation}\label{eq:16''}
\bar x=\frac{Q^2}{2\bar p\cdot q}=\frac{Q^2}{(\bar n\cdot \bar p)(n\cdot q)}\simeq \frac{\bar n\cdot q}{\bar n\cdot \bar p}\,.
\end{equation}
Here $\tau=Q^2/s=x\cdot \bar x$ is the fraction of the energy squared taken by the colliding partons from the protons. The end point corresponds to $\tau\to 1$. As in DIS, we define $\frac{Q}{x}=Q+l^+$, $\frac{Q}{\bar x}=Q+\bar l^-$ with light cone momenta $l^+$ and $\bar l^-$.  The separated scales are
\begin{itemize}
\item hard modes with $q=(Q,Q,0)$ at the hard scale;
\item $n$-collinear modes with $p_c=\left(\frac{Q}{x},\frac{\LQCD^2}{Q},\LQCD\right)\sim (Q+l^+,l^-,\LQCD)$ with invariant mass $p^2\sim\Lambda_{\text{QCD}}^2$;
\item $\bar n$-collinear modes with $\bar p_c=\left(\frac{\LQCD^2}{Q},\frac{Q}{\bar x},\LQCD\right)\sim (\bar l^+,Q+\bar l^-,\LQCD)$ with invariant mass $\bar p^2\sim \Lambda_{\text{QCD}}^2$;
\item soft modes with $p_s\sim(\LQCD,\LQCD,\LQCD)$ at the soft scale.
\end{itemize}
As $x,\bar x\to 1$, the off-shellness of the initial states $Q\frac{(1-x)}{x}\sim l^+$ and $Q\frac{(1-\bar x)}{\bar x}\sim \bar l^-$ goes to $\Lambda_{\text{QCD}}$, bringing in new rapidity singularities arising from the fact that both soft and collinear modes have invariant mass squared of  order $\Lambda^2_{\text{QCD}}$.  These singularities are regulated with the covariant $\eta$ regulator, which allows us to resum the rapidity logarithms by running from $Q$ to $Q\frac{(1-x)}{x}\sim l^+\sim Q\frac{(1-\bar x)}{\bar x}\sim \bar l^-$.

\subsection{Factorization}\label{sec:sec3.2}
A number of papers have discussed factorization of Drell-Yan using SCET \cite{Bauer:2002nz,idilbi05,becher07,chay13,Catani1996,Becher2006b,Idilbi2006}.  Here we follow Ref.\,\cite{Bauer:2002nz}, starting with the unpolarized DY cross section,
\begin{equation}\label{eq:17}
d\sigma=\frac{32\pi^2\alpha^2}{sQ^4}L_{\mu\nu}W^{\mu\nu}\frac{d^3k_1}{(2\pi)^3(2k_1^0)}\frac{d^3k_2}{(2\pi)^3(2k_2^0)}\,,
\end{equation}
where $L_{\mu\nu}$ is the lepton tensor, and $W^{\mu\nu}$ is the DY hadronic tensor.  Equation \req{eq:17} gives
\begin{align}\label{DYdiffcrosssection}
\frac{d\sigma}{dQ^2}=-\frac{2\alpha}{3Q^2s}\int\frac{d^4q}{(2\pi)^3}\delta(q^2-Q^2)\theta(q_0)W(\tau,Q^2)
\end{align}
where $Q^2=\tau s$ is the lepton pair's center-of-mass energy squared.  Summing over final states, we obtain
\begin{equation}\label{eq:19}
W(\tau,Q^2)=-\frac{1}{4}\sum_{\text{spin}}\int d^4y e^{-iq\cdot y}\langle p\bar p|J^{\mu\dag}(y)J_\mu(0)|p\bar p\rangle\,,
\end{equation}
where $J^\mu(y)$ is the QCD current as in Eq.~(\ref{QCDSCETcurrent}).  Near the end-point region, the magnitude of the 3-momentum transferred is
\begin{align}
|\vec q|\leq \frac{Q}{2}(1-\tau),
\end{align}
where $Q=\sqrt{Q^2}$.  As a result, the zero component is
\begin{align}
q_0=Q+\mathcal{O}(1-\tau)\gg|\vec q|.
\end{align}
Therefore the $\delta$-function in \req{DYdiffcrosssection} is expanded,
\begin{align}
\delta(q^2-Q^2)=\frac{1}{2Q}\delta(q_0-Q)+\mathcal{O}(1-\tau)^2.
\end{align}
Carrying out the $q_0$ integration, the hadronic structure function becomes
\begin{align}
W(\tau,Q^2)= -\frac{1}{8Q}\sum_{\rm spins}\int\frac{d^3q}{(2\pi)^3}\int d^4y e^{-iQy_0+i\vec q\cdot \vec y}\langle p\bar p|J^{\mu\dag}(y)J_\mu(0)|p\bar p\rangle.
\end{align}

We match $W$ onto the $\scetii$, and get
\begin{align}
W^{\text{eff}}& =-\oneov{4}\sum_{\sigma,\sigma'}\int\frac{d^3q}{(2\pi)^3}\int d^4y\oneov{2Q}\sum_{\bar w,w}
C^*(Q,Q';\mu_q,\mu)C(\bar w,w,;\mu_q,\mu) \delta_{\bar n\cdot p_n,Q}\delta_{n\cdot\bar p_{\bar n},Q}\label{eq:20}\\
& \times
\langle h(p_n,\sigma)\bar h(\bar p_{\bar n},\sigma')|\bar T[\bar \chi_{\bar n,\bar w'}Y_{\bar n}^\dagger \bar Y_n \gamma_\mu^\perp\chi_{n,w'}(y)]\:T[\bar \chi_{n,w}
\bar Y_{\bar n}^\dagger Y_n \gamma^\mu_\perp\chi_{\bar n,\bar w}(0)]|h(p_n,\sigma)\bar h(\bar p_{\bar n},\sigma')\rangle \,. \nn
\end{align}
Here, we have defined the $\bar n$-direction incoming proton to be carrying momentum $\bar p^\mu=\oneov{2}(n\cdot p_{\bar n}+n\cdot \bar p_r)\bar n^\mu$ with the large component of $\bar p^\mu$ scaling as $n\cdot p_{\bar n}\simeq Q/\bar x \simeq Q$ and the residual momentum $\bar p_r^\mu$ containing the small momentum $\bar p_r^\mu\simeq \bar\ell^-\simeq Q\frac{1-\bar x}{\bar x}$.  Similarly, the $n$-direction incoming proton momentum is $p^\mu=\oneov{2}(\bar n\cdot p_n +\bar n\cdot p_r)n^\mu$ with the large component of $p^\mu$ scaling as $\bar n\cdot p_{n}\simeq Q/x\simeq Q$ and the residual momentum $\bar p_r^\mu$ containing the small momenta $p^\mu_r\simeq \ell^+\simeq Q\frac{1-x}{x}$.
We introduce Kronecker deltas to fix the large components of $p$ and $\bar p$ to be equal to $Q$ and integrate over the residual components of the coordinates in position space.  
The Wilson lines $Y_{\bar n}$ and $\bar Y_n$ are associated with soft radiation from two incoming states,
\begin{align}
Y_n(y)&=P\exp[ig\int_{-\infty}^y ds n\cdot A_{us}(sn)],\nn \\
\bar Y_{\bar n}^\dagger(y)&=P\exp[-ig\int_{-\infty}^y ds\bar n\cdot A_{us}(s\bar n)]\,. \label{eq:23}
\end{align}
The hadronic structure function can be factored into the three sectors,
\begin{align}
W^{\text{eff}}&=-\oneov{4}\sum_\sigma\int\frac{d^3q}{(2\pi)^3}\int \frac{d^4y}{2Q}e^{i\vec q\cdot \vec y}\sum_{\bar w,w}  C^*(Q,Q;\mu_q,\mu)C(w,\bar w;\mu_q,\mu)\nn\\
&\times \langle h(p_n,\sigma)\bar h(\bar p_{\bar n},\sigma)|
\bar T[(\bar \chi_{\bar n,Q}^\alpha)^i
\paren{Y_{\bar n}^\dagger (\gamma_{\mu}^\perp)_{\alpha\beta} \bar Y_n}_{ij}(\chi_{n,Q}^\beta)^j(y)]
\delta_{\bar n\cdot p_n,Q}\nn\\
&\times \delta_{n\cdot\bar p_{\bar n},Q}T[(\bar \chi_{n,w}^\rho)^l\paren{\bar Y_{\bar n}^\dagger(\gamma^\mu_{\perp})_{\rho\lambda} Y_n}_{lm}(\chi_{\bar n,w'}^\lambda)^m(0)]|h(p_n,\sigma)\bar h(\bar p_{\bar n},\sigma)\rangle \label{eq:21}\\
&=-\oneov{4}\sum_\sigma\int \frac{d^4y}{2Q} \delta^3(\vec y) \sum_{w,\bar w}C^*(Q,Q;\mu_q,\mu)C(w,\bar w;\mu_q,\mu)
 \delta_{\bar n\cdot p_n,Q}\delta_{n\cdot \bar p_{\bar n},Q}\nn\\
&\times \langle h(p_n,\sigma)|\bar T[(\bar\chi_{\bar n,Q}^\alpha)^i(y)
(\chi_{\bar n,\bar w}^\lambda)^m(0)]|h(p_n,\sigma)\rangle \nn\\
&\times \langle \bar h(\bar p_{\bar n},\sigma)|T[(\bar \chi_{n,w}^\rho)^l(0)(\chi_{n,Q}^\beta)^j(y)]|\bar h(\bar p_{\bar n},\sigma)\rangle \nn\\
&\times \langle 0 |\bar T[(Y_n^\dagger \bar Y_n)_{ij}(y)]T[(\bar Y_{\bar n}^\dagger Y_n)_{lm}(0)]|0\rangle(\gamma_\mu^\perp)_{\alpha\beta}(\gamma^\mu_\perp)_{\rho\lambda}\,.\label{eq:22}
\end{align}
Integrating over $\vec y$, contracting the color indices and averaging the color of the initial states, we have
\begin{align}
W^{\text{eff}}=&
\frac{1}{4}\int \frac{dy_0}{2Q}\oneov{2}\sum_{w,\bar w}C^*(Q,Q;\mu_q,\mu)C(w,\bar w;\mu_q,\mu) \frac{1}{N_c}\sum_\sigma\langle h(p,\sigma)|\bar \chi_{\bar n,Q}(y_0)\frac{\nslash}{2}\chi_{\bar n,\bar w}(0)|h(p,\sigma)\rangle \delta_{\bar n\cdot p_n,Q}\nn \\
&\times\oneov{N_c}\sum_{\sigma'}\langle \bar h(\bar p_{\bar n},\sigma')|\bar \chi_{n,w}{(0)}\frac{\slashed{\bar n}}{2}\chi_{n,Q}(y_0)|\bar h(\bar p_{\bar n},\sigma')\rangle \delta_{\bar p_{\bar n}\cdot n,Q}\langle 0|\bar T[(Y_{\bar n}^\dagger \bar Y_n)](y_0)T[(\bar Y_{\bar n}^\dagger Y_n)](0)|0\rangle \,.\label{eq:25}
\end{align}
Due to label momentum conservation, $w=Q=\bar w$, and we rewrite the large component of the matter field as $\bar \chi_{n,w}=\bar \chi_n\delta_{w,Q}$.  We insert the identities
\begin{align}\label{fieldshiftid}
\bar\chi_{\bar n,Q}(y_0)&=e^{i\hat\partial_0y_0}\bar\chi_{\bar n,Q}(0)e^{-i\hat\partial_0y_0} \\
\chi_{n,Q}(y_0)&=e^{i\hat\partial_0y_0}\chi_{n,Q}(0)e^{-i\hat\partial_0y_0} 
\end{align}
to shift $\bar\chi_{\bar n}$ and $\chi_n$ to the same spacetime point.  The operator $\hat\partial_0$ is a residual momentum operator that acts on the external states to yield
\begin{align}
\hat\partial_0|\bar h(\bar p_{\bar n},\sigma')\rangle &= \frac{Q}{2}\frac{1-\bar x}{\bar x}|\bar h(\bar p_{\bar n},\sigma')\rangle \\
\langle h(p_n,\sigma)|\hat\partial_0&=\langle h(p_n,\sigma)|\frac{Q}{2}\frac{1-x}{x}\,.
\end{align}
 Thus the hadronic structure function is reduced to
 \begin{align}
 W^{\rm eff}&=
|C(Q;\mu_q;\mu)|^2\delta_{\bar n\cdot p_n,Q}\delta_{n\cdot \bar p_{\bar n},Q}\frac{1}{N_c}\int \frac{dy_0}{2Q}e^{-\frac{i}{2}Q\frac{1-x}{x}y_0}e^{-\frac{i}{2}Q\frac{1-\bar x}{\bar x}y_0}
         \frac{1}{N_c}\langle 0|\bar T[Y_{\bar n}^\dag \bar Y_n](y_0) T[\bar Y_n^\dag Y_{\bar n}](0)|0\rangle 
		 \notag \\
         &\times \frac{1}{2}\sum_\sigma\langle h(p_n,\sigma)|\bar\chi_{\bar n}e^{-i\hat\partial_0y_0}\frac{\slashed{n}}{2}\chi_{\bar n}|h(p_n,\sigma)\rangle
         \frac{1}{2}\sum_{\sigma'}\langle\bar h(\bar p_{\bar n},\sigma')|\bar\chi_{n}e^{i\hat\partial_0y_0}\frac{\slashed{\bar n}}{2}\chi_{n}|\bar h(\bar p_{\bar n},\sigma')\rangle\,.
 \end{align}
As in DIS, we define a hard coefficient $H(Q;\mu)=|C(Q;\mu_q,\mu)|^2$, and two collinear functions 
\begin{align}
\frac{1}{2}\sum_\sigma\delta_{\bar n\cdot p_n,Q}\langle h(p_n,\sigma)|\bar\chi_{\bar n}e^{-i\hat\partial_0y_0}\frac{\slashed{n}}{2}\chi_{\bar n}|h(p_n,\sigma)\rangle
&\equiv \int dr e^{-ir y_0}C_{\bar n}(Q+r;\mu),\\
 \frac{1}{2}\sum_{\sigma'}\delta_{n\cdot \bar p_{\bar n},Q}\langle\bar h(\bar p_{\bar n},\sigma')|\bar\chi_{n}e^{i\hat\partial_0y_0}\frac{\slashed{\bar n}}{2}\chi_{n}|\bar h(\bar p_{\bar n},\sigma')\rangle
&\equiv \int d\bar r e^{i\bar r y_0}C_n(Q+\bar r;\mu)   \,.
\end{align}
The SCET hadronic structure function can then be expressed as
\begin{align}\label{Weffstep2}
W^{\rm eff}=&\frac{H(Q,\mu)}{2QN_c}\int dy_0\,e^{-\frac{i}{2}Q(1-\tau)y_0}\int dr d\bar r \,e^{-iry}e^{-i\bar r y}C_{\bar n}(Q+r;\mu)C_n(Q+\bar r;\mu) 
\nonumber \\ & \times 
\frac{1}{N_c}\langle 0|\bar T[Y_{\bar n}^\dag \bar Y_n](y_0) T[\bar Y_n^\dag Y_{\bar n}](0)|0\rangle ,
\end{align}
where $\mu$ is the arbitrary energy scale brought in by matching QCD onto SCET, and its dependence in the hard coefficient $H(Q;\mu)$ introduced by this matching process is canceled by the dependence in the product of the two collinear functions and one soft function. The collinear functions become collinear factors because momentum conservation forbids collinear radiation into the final state. This then requires an additional rapidity scale $\nu$ to separate soft from collinear modes.  Including the rapidity scale dependence,
\begin{align}
C_{\bar n}(Q+r;\mu) & \to C_{\bar n}(Q+r;\mu,\nu)=\mathcal{Z}_{\bar n}(\mu,\nu)\delta(r)\delta_{\bar n\cdot p_n,Q}, \label{eq:79temp}\\
C_n(Q+\bar r;\mu) & \to C_{n}(Q+\bar r;\mu,\nu)=\mathcal{Z}_n(\mu,\nu)\delta(\bar r)\delta_{n\cdot \bar p_{\bar n},Q}  \,. \label{eq:80temp}
\end{align}
As in DIS, these functions are proportional to $\delta$ functions in $r, \bar r$ because there is no real gluon emission into the final state from either proton.

We redefine the soft Wilson lines analogously to the collinear fields in \req{fieldshiftid}, so that
\begin{align}
\langle 0|\bar T[Y_{\bar n}^\dag \bar Y_n](y) T[\bar Y_n^\dag Y_{\bar n}](0)|0\rangle =\langle 0|\bar T[Y_{\bar n}^\dag \bar Y_n](0)e^{i\hat\partial_0y_0} T[\bar Y_n^\dag Y_{\bar n}](0)|0\rangle \,.
\end{align}
Integrating over $r,\bar r$ in \req{Weffstep2} we obtain
\begin{align}
W^{\rm eff}=&H(Q;\mu)\frac{1}{2QN_c}
\mathcal{Z}_{\bar n}(\mu,\nu)\delta_{\bar n\cdot p_n,Q}
\mathcal{Z}_n(\mu,\nu)\delta_{n\cdot \bar p_{\bar n},Q}
\nonumber \\ &\times
\int dy_0\frac{1}{N_c}\langle 0|\bar T[Y_{\bar n}^\dag \bar Y_n](0)e^{i(\hat\partial_0-\frac{Q}{2}(1-\tau))y_0}T[\bar Y_n^\dag Y_{\bar n}](0)|0\rangle .
\end{align}
We define the DY soft function in momentum space to be
\begin{align}
S^{\rm (DY)}(1-\tau;\mu,\nu)=\frac{1}{N_c}\langle 0|\mathrm{tr}\bar T[Y_{\bar n}^\dag \bar Y_{n}](0)\delta\left(2\hat\partial_0-Q(1-\tau)\right)T[\bar Y_n^\dag Y_{\bar n}](0)|0\rangle   \,. \label{eq:DYsoftfunction}
\end{align}
The hadronic structure function becomes
\begin{align}
W^{\rm eff}=\frac{2\pi}{QN_c}H(Q;\mu)
\mathcal{Z}_{\bar n}(\mu,\nu)\delta_{\bar n\cdot p_n,Q}\mathcal{Z}_n(\mu,\nu)\delta_{n\cdot \bar p_{\bar n},Q}
S^{\rm (DY)}(1-\tau;\mu,\nu),
\end{align}
and the differential cross section is
\begin{align}
\left(\frac{d\sigma}{dQ^2}\right)_{\text{eff}}=\frac{2\alpha^2}{3Q^2s}\frac{2\pi}{N_c}H(Q;\mu)
\mathcal{Z}_{\bar n}(\mu,\nu)\delta_{\bar n\cdot p_n,Q}\mathcal{Z}_n(\mu,\nu)\delta_{n\cdot \bar p_{\bar n},Q}
\frac{1}{Q}S^{\rm (DY)}(1-\tau;\mu,\nu).
\end{align}

The soft function and the collinear functions run to the common rapidity scale $\nu$ in the end-point region, suggesting the soft radiation contains information from both incoming protons.  Since the $n$-direction and $\bar n$-direction collinear functions are each connected to this soft function at low momenta by the rapidity scale $\nu$, they are coupled to each other through the soft radiation. Therefore, in the end-point region, it does not suffice to identify the PDF of each proton with just the $n$- and $\bar n$-collinear functions.

We introduce a luminosity function that defines the $n$-collinear, $\bar n$-collinear and soft functions all together,
\begin{eqnarray}
\label{DYphi}
\mathbf{L}^{n{\bar n}s}(1-\tau;\mu) =  \delta_{\bar n\cdot p_n,Q} {\cal Z}_{n}(\mu,\nu)\, \delta_{n\cdot \bar p_{\bar n},Q}\mathcal{Z}_{\bar n}(\mu,\nu) S^{\rm (DY)}(1-\tau;\mu,\nu)\,.
\end{eqnarray}
On the right-hand side, the $\nu$ dependence of the $n$-collinear, $\bar n$-collinear and soft functions cancels between the three factors.  In order to relate the Drell-Yan luminosity function in \req{DYphi} to the definition of the PDF in DIS, we can express $\mathbf{L}^{n{\bar n}s}$ as
\begin{align}\label{DYluminosity}
\mathbf{L}^{n\bar ns}(1-\tau;\mu) 
&=\int dx d\bar x\:f^{\bar ns}_{q}(\frac{1-x}{x};\mu)f^{ns}_{q'}(\frac{1-\bar x}{\bar x};\mu)I^{\rm (DY)}_{\tau\to1}(1-\tau-\frac{1-x}{x}-\frac{1-\bar x}{\bar x};\mu),
\end{align}
where the two PDFs are defined  in Eq.~(\ref{eq:016}), and $I^{\rm (DY)}_{\tau\to1}(1-\tau;\mu)$
is an interference factor, independent of $\nu$, which represents the effect of the two protons interfering with each other at the DY end point.

With this interference function, the  $\scetii$ hadronic structure function is
\begin{eqnarray}
\label{DYhadtenSCET2}
W^{\rm eff} &=& \frac{2\pi}{QN_c}H( Q ;\mu)   \\
&& \times  \int dxd\bar x  \,f^{\bar ns}_{q}(\frac{1-x}{x};\mu)f^{ns}_{q'}(\frac{1-\bar x}{\bar x};\mu)I^{\rm (DY)}_{\tau\to1}(1-\tau-\frac{1-x}{x}-\frac{1-\bar x}{\bar x};\mu)\,.\nn
 \end{eqnarray}


\subsection{Renormalization and resummation with rapidity}\label{sec:sec3.3}

In this section we study the renormalization at one loop of the collinear and soft functions appearing in the Drell-Yan hadronic structure function in the end-point region, $C_n(Q+\bar r;\mu,\nu)$, $C_{\bar n}(Q+ r;\mu,\nu)$ and $S(1-\tau;\mu,\nu)$. As in DIS, we use the $\eta$ regulator to render rapidity divergences finite.

\subsubsection{Collinear and soft functions to $\mathcal{O}(\alpha_s)$ for DY}
It is easy to show that the collinear functions in DIS and DY are equal. As in DIS, Fig.\ref{fig:1} shows the $\mathcal{O}(\alpha_s^0)$ Feynman diagram for the collinear function. The $n$-direction collinear function tree-level structure calculated from that diagram is
\begin{equation}
C_n^{\rm DY}(Q+\bar r)^{(0)}=C_n^{DIS(0)}=\delta_{\bar n\cdot p_n,Q}\delta(\bar r)m_n^{(0)} \,,\label{eq:31}
\end{equation}
where $m_n$ is 
\begin{equation}
m_n^{(0)}=\oneov{2}\sum_\sigma\bar\xi_n^\sigma \frac{\bnslash}{2}\xi_n^\sigma\,. \label{eq:32}
\end{equation}
The $\bar n$-direction collinear function at leading order is
\begin{equation}
C_{\bar n}^{\rm DY}(Q+\bar r)^{(0)}=\delta_{n\cdot \bar p_{\bar n},Q}\delta(r)m_{\bar n}^{(0)}, \label{eq:33}
\end{equation}
where
\begin{equation}
m_{\bar n}^{(0)}=\oneov{2}\sum_\sigma\bar\xi_{\bar n}^\sigma\frac{\nslash}{2}\xi_{\bar n}^\sigma\,. \label{eq:34}
\end{equation}
The $\mathcal{O}(\alpha_s)$ $n$-collinear function Feynman diagrams are shown in Fig.~\ref{fig:2}. As discussed in the DIS section, Fig.~\ref{fig:2}(a) is the one-loop virtual correction to the collinear function, while Figs.~\ref{fig:2}(b) and \ref{fig:2}(c) are real corrections. We add the diagrams of Figs.~\ref{fig:2}(a) and \ref{fig:2}(b) with the mirror diagrams, and multiplies this by the quark wave function renormalization to obtain
\begin{align}
C_n^{\rm DY}(Q+\bar r;\mu,\nu)^{(1)}&=C_n^{DIS(1)}\nn\\
&=
C_n^{(0)}(Q+\bar r;\mu,\nu) \, \frac{\alpha_s C_F}{\pi}w^2 
\bigg\{ \frac{e^{\epsilon \gamma_E} \Gamma(\epsilon)}{\eta}\bigg(\frac{\mu^2}{m^2_g}\bigg)^\epsilon +\frac{1}{\epsilon}\bigg[ \frac{3}{4} + \ln\frac{\nu}{\bn\cdot p} \bigg] \nn\\
&+\ln\frac{\mu^2}{m_g^2}\ln\frac{\nu}{\bar n\cdot p}
+\frac{3}{4}\ln\frac{\mu^2}{m_g^2}
+\frac{3}{4}-\frac{\pi^2}{6}\bigg\}\,.
\label{eq:41}
\end{align}
For the $\mathcal{O}(\alpha_s)$ $\bar n$-collinear function, we repeat the whole procedure and get
\begin{align}
C_\bn^{\rm DY}(Q+r;\mu,\nu)^{(1)}
&=
C_n^{(0)}(Q+r;\mu,\nu) \, \frac{\alpha_s C_F}{\pi}w^2 
\bigg\{ \frac{e^{\epsilon \gamma_E} \Gamma(\epsilon)}{\eta}\bigg(\frac{\mu^2}{m^2_g}\bigg)^\epsilon +\frac{1}{\epsilon}\bigg[ \frac{3}{4} + \ln\frac{\nu}{n\cdot \bar p} \bigg] \nn\\
&+\ln\frac{\mu^2}{m_g^2}\ln\frac{\nu}{n\cdot \bar p}
+\frac{3}{4}\ln\frac{\mu^2}{m_g^2}
+\frac{3}{4}-\frac{\pi^2}{6}\bigg\}\,.
\label{eq:42}
\end{align}

Next we turn our attention to the soft function. 
The tree-level result is trivial,
\begin{equation}
S(1-\tau)^{(0)}=\frac{\delta(1-\tau)}{Q} \,. \label{eq:47}
\end{equation}
The $\mathcal{O}(\alpha_s)$ soft function Feynman diagrams are shown in Fig.\,\ref{fig:DISsoft} (mirror diagrams are not shown).  The soft Wilson lines in Eq.~(\ref{eq:25}) are defined in \eqref{eq:23}.  Comparing these to the soft Wilson lines in DIS in Eq.\,(\ref{eq:13'}), we find that the $\bar n$-direction gluons are changed from outgoing to incoming. Reference \, \cite{Kang2015} however shows that up to $\mathcal{O}(\alpha_s^2)$, the dijet hemisphere soft function in DIS and DY are equal, so the virtual DY soft function at $\mathcal{O}(\alpha_s)$ is the same as in DIS,
\begin{align}
S_v^{\rm DY} =\delta(1-\tau)\frac{2\alpha_sC_F}{\pi Q}w^2\bigl[-\frac{e^{\epsilon\gamma_E}\Gamma(\epsilon)}{\eta}\paren{\frac{\mu}{m_g}}^{2\epsilon}+\oneov{2\epsilon^2}
+\oneov{\epsilon}\ln\frac{\mu}{\nu}+\ln^2\frac{\mu}{m_g}-\ln\frac{\mu^2}{m_g^2}\ln\frac{\nu}{m_g}-\frac{\pi^2}{24}\bigl] \,.
\label{eq:44'}
\end{align}

The naive contribution to the $\mathcal{O}(\alpha_s)$ real DY soft function shown in Fig. 3(b) is
\begin{align}
\tilde S_r^{\rm DY}=&-4C_Fg^2\mu^{2\epsilon}\nu^\eta \int\frac{d^Dk}{(2\pi)^{D-1}}\delta(k^2-m_g^2)\delta(\ell_0-(k^++k^-))\frac{|2k^3|^{-\eta}}{k^+k^-} \label{eq:48}\\
=&\paren{-\frac{\alpha_sC_F}{2\pi Q}}\left\{2\paren{\ln\frac{m_g^2}{Q^2}}\paren{\oneov{1-\tau}}_+-4\paren{\frac{\ln 1-\tau}{1-\tau}}_+ -\paren{\frac{1}{2}\ln^2\frac{Q^2}{m_g^2}}\delta(1-\tau)\right\}\,,
\label{naivesoftDY}
\end{align}
where $\ell_0 = Q (1-\tau)$.  The measurement $\delta$-function at the end-point region of Drell-Yan process requires the soft momentum $\ell$ to be the symmetric sum of $n$ and $\bar n$ gluon momenta, $\ell_0=k^++k^-$, which has the consequence that there are neither rapidity divergences nor ultraviolet divergences.  In Appendix B, we show that the kinematic constraints in DY imply that no collinear modes overlap with the soft momentum region. However, applying the zero-bin subtraction prescription we used in DIS would require both an $n$-collinear and an $\bn$-collinear subtraction, while the prescription in Ref.~\cite{Hoang:2015iva} has no collinear zero bin in the DY soft function. Thus the approach of Ref.~\cite{Hoang:2015iva} is consistent while our approach is not. 
Thus,
\begin{equation}
S_r^{\text{DY}}=2\tilde S_{r}^{\text{DY}}\,.
\end{equation}

The $\mathcal{O}(\alpha_s)$ expression of the soft function is given by adding virtual and soft pieces with their mirror amplitudes,
\begin{align}
S(1-\tau;\mu,\nu)^{(1)}&=\frac{\alpha_sC_F}{\pi Q}w^2\Bigg[
\Bigg( -\frac{2\Gamma(\epsilon)e^{\gamma_{E}}}{\eta}\paren{\frac{\mu}{m_g}}^{2\epsilon}+\frac{1}{\epsilon^2}+\frac{2}{\epsilon}\ln\frac{\mu}{\nu} \nonumber \\
&+2\ln^2\frac{\mu}{m_g}-2\ln\frac{\mu^2}{m_g^2}\ln\frac{\nu}{m_g}
-\frac{\pi^2}{12}+\frac{1}{2}\ln^2\frac{m_g^2}{Q^2}
\Bigg)\delta(1-\tau)\nonumber \\
&+4\paren{\frac{\ln 1-\tau}{1-\tau}}_+ -2\paren{\ln \frac{m_g^2}{Q^2}}\paren{\oneov{1-\tau}}_+ \Bigg]\,.
\label{eq:53}
\end{align}
Comparing this result with the $\mathcal{O}(\alpha_s)$ $n$- and $\bn$-collinear functions given in Eqs.~\req{eq:41} and \req{eq:42}, we see that the $\nu$-dependence cancels in the cross section at $\mathcal{O}(\alpha_s)$.  Forming the ratio of the DY soft function to the product of the $n$ and $\bn$ DIS soft functions gives an interference factor \req{DYluminosity} that is independent of $\nu$ to this order (and presumably to all orders).

\subsubsection{Anomalous dimensions for collinear and soft functions}
The divergences in UV and rapidity in the collinear and soft functions Eqs.~(\ref{eq:41}), (\ref{eq:42}) and (\ref{eq:53}) can be subtracted by counterterms in textbook fashion. We define the relations between the renormalized and the bare functions as
\begin{align}
C_n(Q+\bar r)^R&=Z_n^{-1}C_n(Q+\bar r)^B, \nn \\
C_{\bar n}(Q+r)^R&=Z_{\bar n}^{-1}C_{\bar n}(Q+r)^B\nn, \\
S(1-\tau)^R&=-\int d\tau' Z_s(\tau'-\tau)^{-1}S(1-\tau')^B \,.\label{eq:54}
\end{align}
Thus, Eqs.\,\eqref{eq:41},~\eqref{eq:42} and \eqref{eq:53} yield for the $\mathcal{O}(\alpha_s)$ collinear and soft renormalization factors 
\begin{align}
Z_n&=1+\frac{\alpha_sC_F}{\pi}w^2\left[\frac{\Gamma(\epsilon)}{\eta}\paren{\frac{\mu}{m_g}}^{2\epsilon}+\oneov{\epsilon}\paren{\frac{3}{4}+\ln\frac{\nu}{\bar n\cdot p}}\right],\label{eq:56}\\
Z_{\bar n}&=1+\frac{\alpha_sC_F}{\pi}w^2\left[\frac{\Gamma(\epsilon)}{\eta}\paren{\frac{\mu}{m_g}}^{2\epsilon}+\oneov{\epsilon}\paren{\frac{3}{4}+\ln\frac{\nu}{n\cdot p}}\right],\label{eq:57}\\
Z_s&=\delta(1-\tau)+\frac{\alpha_sC_F}{\pi}w^2\left[-\frac{2\Gamma(\epsilon)}{\eta}\paren{\frac{\mu}{m_g}}^{2\epsilon}+\oneov{\epsilon^2}+\frac{2}{\epsilon}\ln\frac{\nu}{\mu}\right]\delta(1-\tau)   \,.\label{eq:58}
\end{align}
These obey the consistency condition for Drell-Yan at the limits $x,\bar x\to 1$ and hence $\tau\to 1$,
\begin{equation}\label{eq:59}
Z_H \delta(1-\tau)=Z_{\bar n}^{-1}Z_n^{-1}Z_s^{-1},
\end{equation}
where $Z_H$ is given in Eq.\,\eqref{ZH}.
The logarithms in the collinear function are minimized by setting $\nu_c\sim Q$, while in the soft function $\nu_s\sim\mu\sim\LQCD$.  Therefore we must resume these logarithms both in $\mu$ and $\nu$.
From Eq.~(\ref{eq:56}) to Eq.~(\ref{eq:58}) we also can extract the $\mathcal{O}(\alpha_s)$ anomalous dimensions. The $\mu$ anomalous dimensions are
\begin{align}
\gamma_n^\mu(\mu,\nu_c)&=\frac{2\alpha_sC_F}{\pi}\paren{\frac{3}{4}+\ln\frac{\nu_c}{\bar n\cdot p}},\nn\\
\gamma_{\bar n}^\mu(\mu,\nu_c)&=\frac{2\alpha_sC_F}{\pi}\paren{\frac{3}{4}+\ln\frac{\nu_c}{n\cdot \bar p}},\nn\\
\gamma_s^\mu(\mu,\nu_s)&=\frac{4\alpha_sC_F}{\pi}\ln\frac{\mu}{\nu_s}\delta(1-\tau)\,.
\label{eq:61}
\end{align}
As in DIS, the sum $\gamma_n^\mu(\mu,\nu)\delta(1-\tau)+\gamma_{\bar n}^\mu(\mu,\nu)\delta(1-\tau)+\gamma_s^\mu(\mu,\nu)$ is independent of the rapidity scale $\nu$, as expected.  However, the sum contains a large logarithm of $(\bar n\cdot p) (n\cdot \bar p)\sim Q^2$. The $\nu$ anomalous dimensions are 
\begin{align} 
\gamma_n^\nu(\mu,\nu_c)&=\frac{\alpha_sC_F}{\pi}\ln\frac{\mu^2}{m_g^2},\nn \\
\gamma_{\bar n}^\nu(\mu,\nu_c)&=\frac{\alpha_sC_F}{\pi}\ln\frac{\mu^2}{m_g^2},\nn\\
\gamma_s^\nu(\mu,\nu_s)&=-\frac{2\alpha_sC_F}{\pi}\ln\frac{\mu^2}{m_g^2}\delta(1-\tau)\,.
\label{eq:62}
\end{align}
Unsurprisingly, $\gamma_n^\nu(\mu,\nu)\delta(1-\tau) + \gamma_{\bar n}^\nu(\mu,\nu) \delta(1-\tau)+\gamma_s^\nu(\mu,\nu)=0$ when $\nu_c=\nu_s$ in the limits $x,\bar x\to 1$ and $\tau\to 1$. The presence of $m_g$ suggests the same IR sensitivity as occurred in DIS. As we see in the next section, this IR dependence in anomalous dimensions also shows up in the delta regulator scheme for the divergences in the end-point region.

From the $\mu$ anomalous dimensions in Eq.\eqref{eq:61}, we can see the soft function runs to the scale $\nu$ common also to the collinear functions as we have already seen in the DIS case.  This is problematic because it means the two collinear functions, which are traditionally identified with the proton PDFs, are not independent from each other.  At moderate $x$, these scales would not run to the same point and the two collinear functions can be separated.  Thus at large $x$ the two collinear functions cannot be separated and we do not have a unique way to define independent (and so universal) PDFs for the colliding protons. Preserving the conventional description of the colliding protons in terms of $n$-collinear, $\bar n$-collinear and soft functions, we arrive at the luminosity function in Eq.~\eqref{DYphi}, and at large $x$ the two collinear pieces and one soft piece are related by the common rapidity scale $\nu$. 

Now we connect the running and the resummation results in DY with those in DIS by solving the renormalization equation of the interference factor $I^{\text{DY}}$ we defined in Eq.\,\eqref{DYluminosity}. Using the newly introduced PDF definition in Eq.\,\eqref{eq:011}, we can write the PDF for the $n$-direction incoming proton as
\begin{align}\label{DYnPDF}
f_q^{ns}(\frac{1-x}{x};\mu)&=\frac{\delta(1-x)}{Q} \nn \\
&+\frac{\alpha_s C_F}{\pi Q}\left( \left[\ln\frac{\mu^2}{m_g^2}\ln\frac{\nu_n}{\nu_s}+\frac{3}{4}\ln\frac{\mu^2}{m_g^2}+\frac{3}{4}-\frac{\pi^2}{6}\right]\delta(1-x)+\ln\frac{\mu^2}{m_g^2}\left(\frac{1}{1-x}\right)_+\right)\,,
\end{align}
where $\nu_n$ is the $n$-direction incoming proton near end-point rapidity scale.  Changing $\nu_n$ to $\nu_{\bar n}$ and $x$ to $\bar x$, we have the $\bar n$-direction incoming proton PDF
\begin{align}\label{DYbarnPDF}
f_{q'}^{ns}(\frac{1-\bar x}{\bar x};\mu)&=\frac{\delta(1-\bar x)}{Q} \nn\\
&+\frac{\alpha_s C_F}{\pi Q}\left( \left[\ln\frac{\mu^2}{m_g^2}\ln\frac{\nu_{\bar n}}{\nu_s}+\frac{3}{4}\ln\frac{\mu^2}{m_g^2}+\frac{3}{4}-\frac{\pi^2}{6}\right]\delta(1-\bar x)+\ln\frac{\mu^2}{m_g^2}\left(\frac{1}{1-\bar x}\right)_+\right).
\end{align}
Expanding the interference factor in \req{DYluminosity} in powers of $\alpha_s$,
\begin{align}\label{IDYexpansion}
I^{(DY)}_{\tau\to 1}(1-\tau;\mu)=I^{(DY)}_0+I^{(DY)}_1+...
\end{align}
Plugging Eqs.\,\eqref{DYnPDF}-\eqref{IDYexpansion} into \req{DYluminosity}, we extract
\begin{align}\label{IDYzeroth}
I^{(DY)}_0=2Q\delta(1-\tau),
\end{align}
and the unrenormalized order-$\alpha_s$ interference function
\begin{align}
I^{(DY)}_1(1-\tau;\mu)&=2Q\frac{\alpha_s C_F}{\pi}\Bigg(\bigg[ \oneov{\epsilon^2}+\oneov{\epsilon}\ln\frac{\mu^2}{Q^2}+2\ln^2\frac{\mu}{Q}-\frac{\pi^2}{12}\bigg]\delta(1-\tau)
\notag \\ &
+4\paren{\frac{\ln 1-\tau}{1-\tau}}_+ -2\paren{\ln\frac{\mu^2}{Q^2}}\paren{\oneov{1-\tau}}_+ -\frac{2}{\epsilon}\paren{\oneov{1-\tau}}_+ \bigg), \label{eq:109'}
\end{align}
which is independent of rapidity scale $\nu$.  This result is independent of any infrared scales and is consistent with the DY soft function defined in Ref.\,\cite{becher07}, Eq.\,(45).  The counterterm is
\begin{equation}\label{eq:109''}
Z_I^{\text{DY}}=\delta(1-\tau)+\frac{\alpha_s C_F}{\pi}\bigg\{\left[\oneov{\epsilon^2}+\oneov{\epsilon}\ln\frac{\mu^2}{Q^2}\right]\delta(1-\tau)-\frac{2}{\epsilon}\paren{\oneov{1-\tau}}_+\bigg\},
\end{equation}
and the $\mu$ anomalous dimension is
\begin{equation}\label{eq:109'''}
\gamma_I^{\text{DY}}(1-\tau;\mu)=\frac{4\alpha_s C_F}{\pi }\left[\ln\frac{\mu}{Q}\delta(1-\tau)-\paren{\oneov{1-\tau}}_+\right],
\end{equation}
through which we can resum the logarithms brought in by the interference effect between the two protons.  This anomalous dimension is consistent with Eq.\,(43) of Ref.\,\cite{becher07}.  Note the appearance of the cusp in $\gamma^{\rm DY}_I$, which resums Sudakov double logarithms.  To $\mathcal{O}(\alpha_s)$, the renormalized interference factor is
\begin{align}\label{IDYrenorm}
I^{(DY)}(1-\tau;\mu)&=2Q\delta(1-\tau)+2Q\frac{\alpha_s C_F}{\pi}\bigg(\left[\frac{1}{2}\ln^2\frac{\mu^2}{Q^2}-\frac{\pi^2}{12}\right]\delta(1-\tau) \nn \\
&+4\left(\frac{\ln(1-\tau)}{1-\tau}\right)_+ +2\ln\frac{Q^2}{\mu^2}\left(\frac{1}{1-\tau}\right)_+ \bigg).
\end{align}

\subsubsection{Comparing to perturbative QCD results}
The hard function $H^{\text{DY}}$ we extract from Ref.\,\cite{becher07} is
\begin{equation}\label{eq:100b}
H^{\text{DY}}(Q,\mu)=1+\frac{\alpha_sC_F}{\pi}\left(-\frac12\ln^2\frac{\mu^2}{Q^2}-\frac32\ln\frac{\mu^2}{Q^2}-4+\frac{7\pi^2}{12}\right)  \,.
\end{equation}
Taking $N_c=3$, and inserting the Drell-Yan collinear and soft functions with Eq.\,\eqref{eq:100b} into Eq.\,\eqref{eq:31}, we find at $\mathcal{O}(\alpha_s)$ SCET the Drell-Yan cross section is
\begin{align}
\left(\frac{d\sigma}{dQ^2}\right)_{\text{eff}}&=m_0^2\delta_{n\cdot\bar p_{\bar n},Q}\delta_{\bar n\cdot p_n,Q}\left(\frac{4\pi \alpha^2}{9Q^4}\right)\frac{\alpha_sC_F}{\pi}\bigg\{\left[\frac32\ln\frac{Q^2}{m_g^2}-\frac{5}{2}-\frac{\pi^2}{6}\right]\delta(1-\tau) \label{eq:100c}\\
& \hspace{2.5cm}+4\left(\frac{\ln(1-\tau)}{(1-\tau)}\right)_++2\ln\frac{Q^2}{m_g^2}\left(\frac{1}{1-\tau}\right)_+\bigg\}\nn.
\end{align}
To $\mathcal{O}(\alpha_s)$ in QCD, the quark contribution to the DY cross section is \cite{Field1989}
\begin{align}
\frac{d\sigma}{dQ^2}&=m_0^2\delta_{n\cdot\bar p_{\bar n},Q}\delta_{\bar n\cdot p_n,Q}\frac{4\pi}{9}\frac{\alpha^2}{Q^4}\int_\tau^1\frac{dx_a}{x_a}\int_{\tau/x_a}^1\frac{dx_b}{x_b}\bigg\{G_{p\to q}^{(0)}(x_a)G_{p\to q}^{(0)}(x_b)\nn\\
&\times\bigg(\frac{\sigma_{\text{tot}}^{\text{DY}}}{\sigma_0}\delta(1-z)+\frac{\alpha_s}{\pi}P_{q\to qg}(z)\ln\frac{Q^2}{m_g^2}+2\alpha_s f^{q\text{ DY}}(z)\bigg)\bigg\}\,,\label{eq:100d}
\end{align}
where $z=\tau/(x_ax_b)$, $G_{p\to q}^{(0)}(x_a), G_{p\to q}^{(0)}(x_b)$ are zero-order PDFs, and
\begin{align}
P_{q\to qg}(\tau)&=C_F\left(\frac{1+z^2}{(1-z)_+}+\frac{3}{2}\delta(1-z)\right)\nn\\
\alpha_s f^{q\text{ DY}}(\tau)&=\frac{\alpha_sC_F}{\pi}\bigg\{(1+z^2)\left(\frac{\ln(1-z)}{1-z}\right)_+-\left(\frac{1+z^2}{1-z}\right)\ln z\nn\\
&-(1-z)-\frac{\pi^2}{3}\delta(1-z)
\bigg\}\,,\nn\\
\frac{\sigma_{\text{tot}}}{\sigma_0}&=1+\left(\frac{8\pi}{9}-\frac{7}{3\pi}\right)\alpha_s+\ldots
\label{eq:100e}
\end{align}
In the end point, $z\to 1$ the perturbative QCD Drell-Yan cross section at $\mathcal{O}(\alpha_s)$ becomes
\begin{align}
\frac{d\sigma}{dQ^2}&=m_0^2\delta_{n\cdot\bar p_{\bar n},Q}\delta_{\bar n\cdot p_n,Q}\left(\frac{4\pi\alpha^2}{9Q^4}\right)\frac{\alpha_sC_F}{\pi}
\bigg\{\left[\frac32 \ln\frac{Q^2}{m_g^2}-\frac{7}{4}\right]\delta(1-\tau)+4  \nn \\ \label{eq:100g}
& \hspace{2.5cm}+4\left(\frac{\ln(1-\tau)}{1-\tau}\right)_+ + 2 \ln\frac{Q^2}{m_g^2} \left(\frac{1}{1-\tau}\right)_+\bigg\} \,.
\end{align}
Comparing Eqs.\eqref{eq:100g} and \eqref{eq:100c}, we arrive at the same conclusion as for DIS, that the $\text{SCET}_{\text{II}}$ hadronic structure function reproduces all the low energy physics of the perturbative QCD results in the end-point region up to constant coefficients of  $\delta(1-\tau)$, which is regularization scheme dependent.  As in DIS, this discrepancy is expected since the SCET and QCD calculations use different regularization schemes.

\section{DIS and DY at end point with Delta Regulator}\label{sec:sec4}

The method of the delta regulator was introduced to implement a proper zero-bin subtraction sector so as to remove the overlap between the collinear and soft functions and restore the SCET factorization theorem. In this sense it serves a similar role as the $\eta$-rapidity regulator, except that the latter is gauge invariant and associated with a rapidity scale making resummation in the rapidity region possible. To exhibit the origin of this fact, we repeat our calculations in the previous two sections using the delta regulator, and note the pros and cons of these two regularization schemes line by line.

\subsection{Wilson lines and factorization with delta regulator}

We define the Delta regulator, by adding a constant in the propagator denominators as in Ref.~\cite{Chiu:2009yx},
\begin{equation}\label{eq:63}
\oneov{(p_i+k)^2-m_i^2}\to \oneov{(p_i+k)^2-m_i^2-\Delta_i}\,.
\end{equation}
The subscript $i$ denotes the particle $i$. The form of Eq.~\eqref{eq:63} makes the $\Delta$ regulator behave like a mass shift for the particle $i$. Correspondingly, the collinear Wilson lines are
\begin{align}
W_n&=\sum_{perm}\exp\left[-\frac{g}{\bar n\cdot \mathcal{P}-\delta_1}\bar n\cdot A_n\right],\nn \\
W_{\bar{n}}^\dg&=\sum_{perm}\exp\left[-\frac{g}{n\cdot\mathcal{P}-\delta_2}n\cdot A_{\bar n}\right]\,,\label{eq:64}
\end{align}
while the soft Wilson lines for DIS are
\begin{align}
\tilde{Y}_{\bar{n}}^\dg&=\sum_{perm}\exp\left[-\frac{g}{n\cdot \mathcal{P}_s-\delta_2+i\epsilon}\bar{n}\cdot A_s\right],\nn\\
Y_n&=\sum_{perm}\exp\left[-\frac{g}{\bar n\cdot\mathcal{P}_s-\delta_1-i\epsilon}n\cdot A_s\right]\,,\label{eq:65}
\end{align}
and for DY are
\begin{align}
\tilde{Y}_{\bar{n}}^\dg&=\sum_{perm}\exp\left[-\frac{g}{n\cdot \mathcal{P}_s-\delta_2-i\epsilon}\bar{n}\cdot A_s\right],\nn\\
Y_n&=\sum_{perm}\exp\left[-\frac{g}{\bar n\cdot\mathcal{P}_s-\delta_1-i\epsilon}n\cdot A_s\right]\,,\label{eq:66}
\end{align}
where $\delta_1=\Delta_1/p^+$ and $\delta_2=\Delta_2/p^-$, with $p^+$ or $p^-$ being the collinear momentum in the $n$ or $\bar n$ direction.

Now we repeat the factorization procedure for semi-inclusive DIS and DY using these delta-regulated Wilson lines. Separating the hard collision scale and decoupling soft degrees of freedom from collinear degrees, we reach the same expressions for the $\text{SCET}_{\text{I}}$ hadronic tensor for DIS Eq.~\eqref{eq:05} and for DY Eq.~\eqref{Weffstep2}.  Then we match the DIS and DY hadronic tensors from $\text{SCET}_{\text{I}}$ to $\text{SCET}_{\text{II}}$, and separate soft and collinear modes with an explicit zero-bin subtraction. Adopting all the definitions for the soft function $S(l,\mu)$ in Eq.~\eqref{eq:07}, jet function in Eq.~\eqref{eq:06} and collinear sectors as in Eq.~\eqref{eq:09}, we have the DIS hadronic tensor in $\text{SCET}_{\text{II}}$ with the delta regulator
\begin{align}
(W_{\mu\nu})_{\delta-DIS}^{\rm eff}&=-g_\perp^{\mu\nu}H(Q,\mu)\int _\phi d\ell J_{\bar n}(r;\mu)S(\ell;\mu;\delta_2,m_g^2)C_n(Q-r-\ell;\mu;\delta_2,m_g^2)\,. \label{eq:67}
\end{align}
Likewise with the soft function $S(\ell^+,\bar \ell^-;\mu,\nu)$ in Eq.~\eqref{eq:DYsoftfunction} and two collinear functions in Eqs.~\eqref{eq:79temp} and \eqref{eq:80temp}, the DY hadronic structure function in $\text{SCET}_{\text{II}}$ with delta regulator is
\begin{align}
(W)_{\delta-DY}^{eff}&=\frac{2\pi}{QN_c}H(Q;\mu_q,\mu)C_n(Q;\mu;\delta_2,m_g^2)C_{\bar n}(Q;\mu;\delta_1,m_g^2)\frac{1}{Q}S(1-\tau;\mu;\delta_1,\delta_2,m_g^2)\,.\label{eq:68}
\end{align}
The notation $\phi$ on the integral emphasizes the need to remove the overlap of the zero bins of each function.

\subsection{Renormalization and running with the delta regulator}
\subsubsection{DIS collinear and soft functions}
For DIS, the naive virtual $n$-collinear function shown in Fig.\ref{fig:2}(a) is
\begin{align}
\tilde{C}_n^v=&(2ig^2C_F)\delta(k^-)\mu^{2\epsilon}\int \frac{d^dq}{(2\pi)^d}\frac{1}{-q^-+\delta_1+i\epsilon}\frac{p^-+q^-}{(p^-+q^-)q^+-q_\perp^2-\Delta_2+i\epsilon}\frac{1}{q^-q^+-q_\perp^2-m_q^2+i\epsilon}\nn\\
=&\left(-\frac{\alpha_s C_F}{2\pi}\right)\delta(k^-)\Bigg(\frac{1}{\epsilon}\left(-\ln\frac{\delta_1}{p^-}-1\right)-\ln\frac{\mu^2}{m_g^2}\left(\ln\frac{\delta_1}{p^-}+1\right)\nn\\
&-\left[\ln\left(1-\frac{\Delta_2}{m_g^2}\right)\ln\frac{\Delta_2}{m_g^2}+1-\frac{\Delta_2/m_g^2}{\frac{\Delta_2}{m_g^2}-1}\ln\frac{\Delta_2}{m_g^2}+Li_2\left(\frac{\Delta_2}{m_g^2}\right)-\frac{\pi^2}{6}\right]\Bigg)\,.\label{eq:69}
\end{align}
We see that $\Delta_2$ is the infrared regulator for the quark propagator, which effectively is the quark mass in the loop integral.
The zero-bin amplitude for this virtual function is 
\begin{align}
C_{n}^{v\phi}&=(-2ig^2C_F)\delta(k^-)\mu^{2\epsilon}\int \frac{d^dq}{(2\pi)^d}\frac{1}{q^--\delta_1+i\epsilon}\frac{1}{q^+-\delta_2+i\epsilon}\frac{1}{q^2-m_g^2+i\epsilon}\nn\\
&=\left(\frac{-\alpha_sC_F}{2\pi}\right)\delta(k^-)\Bigg(\frac{1}{\epsilon^2}+\frac{1}{\epsilon}\ln\frac{\mu^2}{\delta_1\delta_2}
\nn \\
&~~~+\ln\left(\frac{\mu^2}{m_g^2}\right)\ln\frac{\mu^2}{\delta_1\delta_2}-\frac{1}{2}\ln^2\frac{\mu^2}{m_g^2}
-Li_2\left(1-\frac{\delta_1\delta_2}{m_g^2}\right)+\frac{\pi^2}{12}\Bigg)\,.\label{eq:70}
\end{align}
For the real collinear function, the naive real collinear amplitudes only get contributions from the soft momentum region, which are their exact zero-bin subtraction amplitudes. Thus, after the zero-bin subtractions, the real collinear function amplitudes shown in Fig.\,\ref{fig:2}(b) and \ref{fig:2}(c) vanish,
\begin{equation}\label{eq:71}
\tilde C_n^r=C_n^{r\phi}\Rightarrow C_n^r=\tilde C_n^r-C_n^{r\phi}=0\,.
\end{equation}
After multiplying the calculated amplitudes in Eqs.~\eqref{eq:69} and \eqref{eq:70} by 2 for their mirror images, we have the collinear function with quark wave function renormalization in semi-inclusive DIS with the delta regulator
\begin{align}
C_n^v=&2(\tilde C_n^v-\tilde C_n^{v\phi})\nn\\
=&2\left(\frac{-\alpha_s C_F}{2\pi}\right)\delta(k^-)
\Bigg(-\frac{1}{\epsilon^2}-\frac{1}{\epsilon}\left(\ln\frac{\mu^2}{\Delta_2}+\frac{3}{4}\right)+\frac{\pi^2}{12}-\frac{3}{4}+\frac12\ln^2\frac{\mu^2}{m_g^2}-\ln\frac{\mu^2}{m_g^2}\left(\ln\frac{\mu^2}{\Delta_2}+\frac{3}{4}\right)
\nn\\ &+Li_2\left(1-\frac{\delta_1\delta_2}{m_g^2}\right)-Li_2\left(\frac{\Delta_2}{m_g^2}\right)
+\ln\frac{\Delta_2}{m_g^2}\left(\frac{\Delta_2/m_g^2}{\frac{\Delta_2}{m_g^2}-1}-\ln\left(1-\frac{\Delta_2}{m_g^2}\right)\right)\Bigg)\,.
\label{eq:72}
\end{align}
The infrared part of the final result of the $n$-collinear function is independent of $\delta_1$, which is the infrared regulator of the $n$-direction Wilson line.  In contrast, using the rapidity $\eta$ regulator exhibited rapidity divergences in the $n$-collinear function in Eq.~\eqref{eq:24temp} brought in by the  $n$-direction Wilson line. 
The naive virtual soft function for DIS shown in Fig.~\ref{fig:DISsoft}(a) is the same as the zero bin of the virtual collinear function, since the momentum contributing to that integral comes from the same soft region
\begin{align}
\tilde S_v&=\left(-\frac{\alpha_sC_F}{2\pi}\right)\delta(l)\bigl\{\frac{1}{\epsilon^2}+\frac{1}{\epsilon}\ln\frac{\mu^2}{\delta_1\delta_2}+\ln\frac{\mu^2}{m_g^2}\ln\frac{\mu^2}{\delta_1\delta_2}-\frac{1}{2}\ln^2\frac{\mu^2}{m_g^2}-Li_2\left(1-\frac{\delta_1\delta_2}{m_g^2}\right)+\frac{\pi^2}{12}\bigl\}.\label{eq:73}
\end{align}
The naive real soft function shown in Fig.~\ref{fig:DISsoft}(b) is
\begin{align}
\tilde S_r=&(4\pi g^2C_F)\mu^{2\epsilon}\int \frac{d^{4-2\epsilon}k}{(2\pi)^{4-2\epsilon}}\delta(k^2-m_g^2)\delta(l-k^-)\theta(k^0)\frac{1}{k^+-\delta_2}\frac{1}{k^--\delta_1}\nn\\
=&\left(\frac{\alpha_sC_F}{2\pi}\right)\frac{1}{Q}
\Bigg(-\frac{1}{\epsilon}\delta(z)\ln\frac{-\delta_1}{Q}+\left(\frac{1}{z}\right)_+\left(\frac{1}{\epsilon}+\ln\frac{-\mu^2}{\delta_2Q}\right)-\delta(z)\ln\left(-\frac{\delta_1}{Q}\right)\ln\frac{-\mu^2}{\delta_2Q}\Bigg)\,,\label{eq:74}
\end{align}
where $zQ =l$, and $z$ is dimensionless.  We omit the term proportional to $\ln(1-z)\left(\frac{1}{z}\right)_+$, which contributes a constant in the end-point limit $z\to 0$.  The delta regulator restricts the integrals leading to Eqs.~\eqref{eq:73} and \eqref{eq:74} to the soft momentum region, so we do not need to subtract the collinear overlap. This differs from the prescription with the $\eta$ regulator, which serves as a smooth step function in the loop integral and may leave residual overlap with the collinear function that must be eliminated by subtracting. Multiplying Eqs.~\eqref{eq:73} and \eqref{eq:74} by 2 for their mirror images, we get the soft function with the delta regulator
\begin{align}
S=&2(\tilde S_v+\tilde S_r)\nn\\
=&2\left(-\frac{\alpha_sC_F}{2\pi Q}\right)\Bigg(+\frac{1}{\epsilon^2}\delta(z)+\frac{1}{\epsilon}\bigl[\delta(z)\ln\frac{\mu^2}{\delta_1\delta_2}+\delta(z)\ln(-\frac{\delta_1}{Q})-\left(\frac{1}{z}\right)_+\bigl]-\left(\frac{1}{z}\right)_+\ln\frac{-\mu^2}{\delta_2Q}\nn\\
&+\ln(-\frac{\delta_1}{Q})\ln\frac{-\mu^2}{\Delta_2}\delta(z)-\frac{\pi^2}{12}\delta(z)
+\ln\frac{\mu^2}{m_g^2}\ln\frac{\mu Q}{\delta_1 m_g}\delta(z)-Li_2\left(1-\frac{\delta_1\delta_2}{m_g^2}\right)\delta(z)\Bigg)\,. \label{eq:75}
\end{align}
Introducing $\kappa$ to make the arguments of the logarithms dimensionless as in Eq.~\eqref{eq:75} and choosing $-\delta_2Q= m_g^2$, we can recombine logarithms to show that the infrared divergence in the soft function is independent of $\delta_1$.  We can make this choice to relate the regulators, because  in the soft function one of the three infrared delta regulators, $\delta_1,\delta_2$ and $m_g^2$ is redundant, and the system is underconstrained.  Again this is very different from what we obtain by using the $\eta$ regulator in Eq.~\eqref{finalsoft}, where we separate rapidity divergences from infrared divergences and get a result containing both rapidity and IR divergences, each with an appropriate regulator, $\eta$ and $m_g^2$.
The counterterms that renormalize the soft and collinear functions in Eqs.~\eqref{eq:75} and \eqref{eq:72} are
\begin{align}
Z_n&=1+\frac{\alpha_sC_F}{\pi}\left[\frac{1}{\epsilon^2}+\frac{1}{\epsilon}\left(\frac{3}{4}+\ln\frac{\mu^2}{\Delta_2}\right)\right]\label{eq:76}\\
Z_s&=\delta(z)-\frac{\alpha_sC_F}{\pi}\left[\frac{1}{\epsilon^2}\delta(z)+\frac{1}{\epsilon}\left[-\left(\frac{1}{z}\right)_+ +\ln\frac{\mu^2}{\delta_1\delta_2}\delta(z)+\delta(z)\ln(-\frac{\delta_1}{Q})\right]\right] \,.\label{eq:77}
\end{align}
The result \req{eq:75} is consistent with perturbative QCD in the end-point limit, as we show later in this section; however, it differs from Eq.~(A.5) of Ref. \cite{Chay:2013zya} which is also performed in the delta-regulator scheme. The last term of Eq.~(A.5) in Ref. \cite{Chay:2013zya} is not shown in the body of the paper, as it  should not be included in the combined result to be consistent with QCD.

To check our results with the DIS consistency condition Eq.\,\eqref{consistencycond}, we must first calculate the counterterm of the jet function with the delta regulator.  The calculation is carried out in Appendix \ref{app:jetfndeltareg}.  The result is
\begin{align}\label{ZJdelta}
Z_J=\delta(z)+\frac{\alpha_s C_F}{2\pi}\left(\frac{1}{\epsilon^2}\delta(z)+\oneov{\epsilon}\left(\delta(z)\left(\frac{3}{4}+\ln\frac{\mu^2}{\Delta_1}+\ln\frac{-\Delta_1}{(n\cdot p)^2}\right)-\left(\frac{1}{z}\right)_+\right)\right)\,.
\end{align}
Combining this with Eqs.~\eqref{eq:76} and \eqref{eq:77}, we verify the consistency condition Eq.~\eqref{consistencycond}.  The anomalous dimensions are
\begin{align}
\gamma^\mu_n=&\frac{2\alpha_s C_F}{\pi}\oneov{\epsilon}\left(\frac{3}{4}+\ln\frac{\mu^2}{\Delta_2}\right) \label{anomdimmunDISdelta} \\
\gamma^\mu_s=&-\frac{2\alpha_s C_F}{\pi}\left(\frac{1}{\epsilon^2}\delta(z)+\frac{1}{\epsilon}\left(-\left(\frac{1}{z}\right)_+ +\delta(z)\ln\frac{\mu^2}{-\Delta_2}\right)\right)\,.   \label{anomdimmusDISdelta}
\end{align}
Analogous to Eqs.\eqref{anomdimmuDIS} and \eqref{anomdimnuDIS}, we can see that 1) because we only treat the rapidity divergences in the semi-inclusive region as one type of infrared divergence, we cannot separate and resum it using the dimensional regularization scale $\mu$.  2)  Similar to the $\eta$ regulator, the sum of the anomalous dimensions $\gamma^\mu=\gamma^{\mu}_n\delta(z)+\gamma^\mu_s$ from Eqs.\,\eqref{anomdimmunDISdelta} and \eqref{anomdimmusDISdelta} is independent of the additional scale $\Delta_2$.  However, the presence of $\Delta_2$ means the running of both the collinear and soft functions  is nonperturbative.  Since the delta regulator and $\eta$ regulator both exhibit nonperturbative running, our calculations suggest that the dependence on the infrared physics is independent of the regulator.  As a consequence, combining the collinear and soft functions into the new definition of the PDF in Eq.\,\eqref{eq:016} is justified as a regulator-independent choice.

With the counterterms given in Eqs.\,\eqref{eq:76} and \eqref{eq:77}, we choose $-\delta_2 Q=m_g^2$, subtract them along with the wave-function renormalization given in Eq.~\eqref{eq:A14} from the collinear function in Eq.~\eqref{eq:72} and soft function in \req{eq:75}, and let $\delta_1\to 0$ to obtain the renormalized collinear and soft functions
\begin{align}
{C}_n^R&=\left(-\frac{\alpha_s C_F}{\pi}\right)\delta(k^-)\left[\frac{\pi^2}{12}-\frac34-\frac12\ln^2\frac{\mu^2}{m_g^2}-\frac34\ln\frac{\mu^2}{m_g^2}\right]\label{eq:134'temp}\\
{S}^R&=\left(-\frac{\alpha_s C_F}{\pi Q}\right)\left[-\left(\frac{1}{z}\right)_+\ln\frac{\mu^2}{m_g^2}-\frac{\pi^2}{4}\delta(z)+\frac12\ln^2\frac{\mu^2}{m_g^2}\delta(z)\right]\,.\label{eq:134''temp}
\end{align}
We insert Eq.~\eqref{eq:134'''temp}, Eq.~\eqref{eq:134''temp}, renormalized final-jet function Eq.~\eqref{eq:A17} and the hard function Eq.~\eqref{eq:50'} into the hadronic tensor Eq.~\eqref{eq:67} and replace $z$ with $(1-x)$ to obtain
\begin{align}
(W_{\mu\nu})_{\delta-\text{DIS}}^{\text{eff}}&=2m_0\delta_{\bar n\cdot\tilde p,Q}\bigg\{\delta(1-x)+\frac{\alpha_s C_F}{\pi}\bigg[\left(-\frac34\ln\frac{m_g^2}{Q^2}\right)\delta(1-x) \nn\\
&-\left(\frac{1}{1-x}\right)_+\bigg(\ln\frac{m_g^2}{Q^2}
+\frac34\bigg)+\left(\frac{\ln(1-x)}{1-x}\right)_++\left(\frac{15}{8}-\frac{\pi^2}{12}\right)\delta(1-x)\bigg]\bigg\}\,.
\label{eq:134'''temp}
\end{align}
Again, we reproduced the perturbative QCD result except for the constant coefficient of the $\delta(1-x)$ term, which depends on the regularization scheme we choose.

\subsubsection{DY collinear and soft functions}
The virtual and real collinear functions of DY are the same as in DIS, with the $\bar n$-collinear function regulated by $\Delta_1$ and the $n$-collinear function regulated by  $\Delta_2$,
\begin{align}
C_n^{\rm DY}=&2\paren{-\frac{\alpha_s C_F}{2\pi}}\delta(k^-)\Bigg(-\oneov{\epsilon^2}-\oneov{\epsilon}\paren{\ln\frac{\mu^2}{\Delta_2}+1}+\frac{\pi^2}{12}-1-\ln\frac{\mu^2}{m_g^2}\paren{\ln\frac{\mu^2}{\Delta_2}+1}+\frac12\ln^2\frac{\mu^2}{m_g^2}\nn\\
&+Li_2\paren{1-\frac{\delta_1\delta_2}{m_g^2}}-Li_2\paren{\frac{\Delta_2}{m_g^2}}
+\ln\frac{\Delta_2}{m_g^2}\paren{\frac{\Delta_2/m_g^2}{\frac{\Delta_2}{m_g^2}-1}-\ln\paren{1-\Delta_2/m_g^2}}\Bigg)\,,\label{eq:78}\\
C_{\bar n}^{\rm DY}=&2\paren{-\frac{\alpha_s C_F}{2\pi}}\delta(k^+)\Bigg(-\oneov{\epsilon^2}-\oneov{\epsilon}\paren{\ln\frac{\mu^2}{\Delta_1}+1}+\frac{\pi^2}{12}-1-\ln\frac{\mu^2}{m_g^2}\paren{\ln\frac{\mu^2}{\Delta_1}+1}+\frac12\ln^2\frac{\mu^2}{m_g^2}
\nn\\&
+Li_2\paren{1-\frac{\delta_1\delta_2}{m_g^2}}-Li_2\paren{\frac{\Delta_1}{m_g^2}}
+\ln\frac{\Delta_1}{m_g^2}\paren{\frac{\Delta_1/m_g^2}{\frac{\Delta_1}{m_g^2}-1}-\ln\paren{1-\frac{\Delta_1}{m_g^2}}}\Bigg)\,.\label{eq:79}
\end{align}
The virtual soft function for DY is also the same as in DIS,
\begin{align}
S_v^{\rm DY}=\,&\,2\!\paren{\!-\frac{\alpha_sC_F}{2\pi}\frac{1}{Q}}\delta(1-\tau)\Bigg[\oneov{\epsilon^2}+\oneov{\epsilon}\ln\frac{\mu^2}{\delta_1\delta_2}
+\ln\frac{\mu^2}{m_g^2}\ln\frac{\mu^2}{\delta_1\delta_2}-\oneov{2}\ln^2\!\!\paren{\frac{\mu^2}{m_g^2}}+\frac{\pi^2}{12}\nn \\
& \hspace{5cm} -Li_2\!\paren{1-\frac{\delta_1\delta_2}{m_g^2}}\!\Bigg]\,.\label{eq:80}
\end{align}
The real piece of the DY soft function is
\begin{align}
S_r^{\rm DY}&=-2(2\pi g^2C_F)\frac{\mu^{2\epsilon}}{(2\pi)^{4-2\epsilon}}\nn\\
& \times \int dk^+ dk^-\int d\Omega_{1-\epsilon}\frac{d(k_\perp^2)}{2}(k_\perp^2)^{-\epsilon}\frac{\delta(k^+k^--k_\perp^2-m_g^2)\delta(\ell_0-(k^++k^-))}{(k^+-\delta_1)(k^--\delta_2)}
\nonumber \\
&=2\left(\frac{\alpha_sC_F}{2\pi}\right)\frac{1}{Q}\Bigg(\left(\frac12\ln^2\frac{Q^2}{m_g^2}\right)\delta(1-\tau)-\left(\frac{1}{1-\tau}\right)_+\ln\left(\frac{m_g^2}{Q^2}\right)+2\left(\frac{\ln (1-\tau)}{1-\tau}\right)_+\Bigg)\,.
\end{align}
We obtain the above result by setting $\delta_1,\delta_2$ to 0, which has the exact form of the real contribution to the soft function in the $\eta$-regulator scheme Eq.\eqref{eq:48}.  This is reasonable because the $\delta_i$ do not regulate any divergences in the integral and the infrared divergence is regulated by $m_g^2$.  Since there is only one infrared divergence, the regulators $\delta_1,\delta_2$ are redundant, similar to the DIS case.
The soft function for DY is
\begin{align}
S^{\rm DY}&=S_v^{\rm DY}+S_r^{\rm DY}\nn\\
&=\left(-\frac{\alpha_s C_F}{\pi}\right)\frac{1}{Q}\Bigg\{ \frac{1}{\epsilon^2}\delta(1-\tau)+\frac{1}{\epsilon}\ln\frac{\mu^2}{\delta_1\delta_2}\delta(1-\tau)-\Bigg[\frac12\ln^2\frac{Q^2}{m_g^2}+\frac{\pi^2}{12}+\frac12\ln^2\frac{\mu^2}{m_g^2}
\nn\\
&-\ln\frac{\mu^2}{m_g^2}\ln\frac{\mu^2}{\delta_1\delta_2}+Li_2\left(1-\frac{\delta_1\delta_2}{m_g^2}\right)\Bigg]\delta(1-\tau)+2\left(\frac{1}{1-\tau}\right)_+\ln\left(\frac{m_g^2}{Q^2}\right)-4\left(\frac{\ln (1-\tau)}{1-\tau}\right)_+\Bigg\}\,.\label{eq:145'}
\end{align}
Therefore, the counterterms for the DY collinear and soft functions are
\begin{align}
Z_n&=1+\frac{\alpha_s C_F}{\pi}\left[\oneov{\epsilon^2}+\oneov{\epsilon}\paren{\frac34+\ln\frac{\mu^2}{\Delta_2}}\right]\,,\nn\\
Z_{\bar n}&=1+\frac{\alpha_s C_F}{\pi}\left[\oneov{\epsilon^2}+\oneov{\epsilon}\paren{\frac34+\ln\frac{\mu^2}{\Delta_1}}\right]\,,\nn\\
Z_s&=\delta(1-\tau)-\frac{\alpha_s C_F}{\pi}\left[\oneov{\epsilon^2}+\oneov{\epsilon}\ln\frac{\mu^2}{\delta_1\delta_2}\right]\delta(1-\tau)\,,\label{eq:83}
\end{align}
which are regulator dependent and satisfy the consistency condition at $x,\bar x\to 1$ and $\tau\to 1$. The anomalous dimensions for the DY collinear and soft functions are
\begin{align}
\gamma_n^\mu&=\frac{2\alpha_sC_F}{\pi}\paren{\frac34+\ln\frac{\mu^2}{\Delta_2}}\nn\\
\gamma_{\bar n}^\mu&=\frac{2\alpha_s C_F}{\pi}\paren{\frac34+\ln\frac{\mu^2}{\Delta_1}}\nn\\
\gamma_s^\mu&=-\frac{2\alpha_s C_F}{\pi}\ln\frac{\mu^2}{\delta_1\delta_2}\delta(1-\tau)\,.\label{eq:84}
\end{align}
The delta regulators cancel in the sum of the anomalous dimensions in the end-point region, and a large logarithm in $(n\cdot p)(\bar n\cdot\bar p)\sim -Q^2$ remains.
Similar to the DIS case, each piece of the collinear and soft functions is dependent on the infrared physics regardless of the regularization scheme.  As a result, combining the soft and two collinear functions to define the new luminosity function as in Eq.\,\eqref{DYphi} is a regulator-independent choice.
The renormalized $n$- and $\bar n$-collinear functions are
\begin{align}
{C}_n^{\text{DY}-R}&=\left(-\frac{\alpha_sC_F}{\pi}\right)\delta(k^-)\bigg[\frac{\pi^2}{12}-\frac34+\frac12\ln^2\frac{\mu^2}{m_g^2}-\ln\frac{\mu^2}{m_g^2}\ln\frac{\mu^2}{\Delta_2}+Li_2\left(1-\frac{\delta_1\delta_2}{m_g^2}\right)
\nn\\&
-Li_2\left(\frac{\Delta_2}{m_g^2}\right)+\ln\frac{\Delta_2}{m_g^2}\left(\frac{\Delta_2/m_g^2}{\frac{\Delta_2}{m_g^2}-1}-\ln\left(1-\frac{\Delta_2}{m_g^2}\right)\right)\bigg]\label{eq:141'temp}\\
{C}_{\bar n}^{\text{DY}-R}&=\left(-\frac{\alpha_s C_F}{\pi}\right)\delta(k^+)\bigg[\frac{\pi^2}{12}-\frac34+\frac12\ln^2\frac{\mu^2}{m_g^2}-\ln\frac{\mu^2}{m_g^2}\ln\frac{\mu^2}{\Delta_1}+Li_2\left(1-\frac{\delta_1\delta_2}{m_g^2}\right)\nn\\
&-Li_2\left(\frac{\Delta_2}{m_g^2}\right)+\ln\frac{\Delta_1}{m_g^2}\left(\frac{\Delta_1/m_g^2}{\frac{\Delta_1}{m_g^2}-1}-\ln\left(1-\frac{\Delta_1}{m_g^2}\right)\right)\bigg]\,.\label{eq:141''temp}
\end{align}
The renormalized soft function is
\begin{align}
{S}^{\text{DY}-R}&=\left(-\frac{\alpha_sC_F}{\pi}\right)\frac{1}{Q}\Bigg(\bigg[\ln\frac{\mu^2}{m_g^2}\ln\frac{\mu^2}{\delta_1\delta_2}-\frac12\ln^2\left(\frac{\mu^2}{m_g^2}\right)+\frac{\pi^2}{12}-Li_2\left(1-\frac{\delta_1\delta_2}{m_g^2}\right)\nn\\
&-\frac12\ln^2\frac{Q^2}{m_g^2}\bigg]\delta(1-\tau)-2\left(\frac{1}{1-\tau}\right)_+\ln\left(\frac{m_g^2}{Q^2}\right)+4\left(\frac{\ln (1-\tau)}{1-\tau}\right)_+\Bigg)\,.\label{eq:141'''temp}
\end{align}
Inserting Eqs.~\eqref{eq:141'temp}-\eqref{eq:141'''temp} with the hard function Eq.~\eqref{eq:100c} into the DY hadronic structure function Eq.~\eqref{eq:68}, we obtain
\begin{align}
(W^{\mu\nu})_{\text{DY}-\delta}^{\text{eff}}&=\left(\frac{\alpha_sC_F}{\pi}\right)\bigg[\left(\frac{2}{1-\tau}\right)_+\ln\frac{Q^2}{m_g^2}+\frac32\ln\frac{Q^2}{m_g^2}\delta(1-\tau)+4\left(\frac{\ln(1-\tau)}{1-\tau}\right)_+\nn\\
&+\left(\frac32-\frac{7\pi^2}{12}\right)\delta(1-\tau)\bigg]\,.\label{eq:141''''temp}
\end{align}
We can clearly see that Eq.\eqref{eq:141''''temp} reproduces the perturbative QCD result up to the constant coefficient of $\delta(1-\tau)$ which is due to the regularization scheme.

We can also compute the interference factor defined in Eq.~\eqref{DYluminosity} with the soft functions in DIS Eq.~\eqref{eq:77} and DY Eq.~\eqref{eq:145'} as
\begin{align}
I^{\rm DY}&=2Q\delta(1-\tau)+2Q\frac{\alpha_s C_F}{\pi}\Bigg\{\left(\frac{1}{\epsilon^2}+\frac{1}{\epsilon}\ln\frac{\mu^2}{Q^2}+\frac{1}{2}\ln^2\frac{\mu^2}{Q^2}-\frac{\pi^2}{12}\right)\delta(1-\tau)
\nn \\
&+\left(-\frac{2}{\epsilon} +2\ln\frac{Q^2}{\mu^2}\right)\left(\frac{1}{1-\tau}\right)_++4\left(\frac{\ln(1-\tau)}{1-\tau}\right)_+\Bigg\}\,.\label{eq:152}
\end{align}
In relating the DY and DIS soft functions, we exploit the redundancy of our IR regulators and set $\delta_1\delta_2=m_g^2$ in the virtual contribution to the DY soft function.  Except for the constant coefficient of $\delta(z)$, we have the exact same interference factor as Eq.~\eqref{eq:109'} obtained using the rapidity regulator.

\section{Conclusions} \label{sec:conclusion}

In this paper, we have studied the deep inelastic scattering and Drell-Yan processes in the end-point $x\to 1$ ($\tau\to1$) region using both the $\eta$-rapidity regulator and the $\delta$ regulator.  In this region, both DIS and DY exhibit a large Sudakov logarithm, arising as the collinear and soft degrees of freedom approach the same invariant mass scale, which becomes much smaller than the collision center-of-mass scale.  Using soft collinear effective theory and the covariant rapidity regulator to separate collinear and soft degrees of freedom, we see this large logarithm as a logarithm of the ratio of collinear and soft rapidity scales.  We had previously resummed this end-point-region rapidity logarithm in DIS using the rapidity renormalization group, and here we additionally showed how the logarithm of rapidity scales corresponds to the well-known threshold logarithm by transforming the result to Mellin space where it is seen as a divergence going as $\ln N$ for $N\gg 1$.  We also confirmed our previous results for DIS by comparing the same calculations in the $\delta$-regulator scheme and verified agreement with the perturbative QCD result in the limit $x\to 1$.  However, it is notable that the $\delta$ regulator does not provide a convenient mechanism to resum the logarithmic enhancements, which have been argued to be operative even well away from the true end point.

Although separating the parton distribution function in the end-point region into collinear and soft factors brings in dependence on an infrared scale, the rapidity factorization is rigorous, as proven by its successfully reproducing the standard results.  Indeed, the factorization cures the problematic large logarithm, which would otherwise spoil the convergence of the effective theory expansion in the threshold region.  From this point of view, rapidity factorization (and summation) is necessary, even if the running must at some point be reabsorbed into the function chosen to model the PDF at the hadronic scale.  We remark that our definition of the PDF smoothly goes over to the traditional definition away from the end point, and we undertake fitting the experimentally determined PDF to our factorized form in a future publication.  The tangible gain from our analysis is that the running in rapidity we identify may help explain the steep falloff in the PDFs near the end point.

We demonstrated that this rapidity factorization works more generally by performing the same analysis on DY processes.  We resummed the single large rapidity logarithm and compared the resulting factorized collinear functions to the definition of the end-point-region PDF we obtained in DIS.  Morevoer, we verified the results by calculating again in the $\delta$-regulator scheme and by comparing to the perturbative QCD result.  The success of the resummation establishes that rapidity factorization of the PDF is valid also in DY processes, and the parton luminosity function can be related to the PDFs measured in DIS.  

An interesting outcome of separating the DY collinear functions into soft and collinear factors is that the soft radiation necessarily couples to both incoming $n$ and $\bn$ protons.  Consequently there is only a single soft function and the $n$ and $\bn$ parton distribution functions can only be exhibited as separate factors by defining an interference factor.  The hadronic structure function in $\scetii$ has the form
\begin{eqnarray}
W^{\rm eff} =\frac{2\pi}{QN_c}H( Q ;\mu)   
 \int dx d\bar x  \,f_q^{\bar ns}(\frac{1-x}{x};\mu)f_{q'}^{ns}(\frac{1-\bar x}{\bar x};\mu)I^{\rm (DY)}_{\tau\to1}(1-\tau;\mu)\,,~~
 \end{eqnarray}
in which each $\phi(q;m)$ is a PDF defined to be identical to the PDF determined from DIS in the end-point region, and $I^{\rm (DY)}_{\tau\to1}(1-\tau;\mu)$ is the interference factor, whose renormalized form is given in \req{IDYrenorm}.  Calculating its running proves that $I^{\rm (DY)}_{\tau\to 1}$ is a nontrivial function and is independent of the rapidity scale.   The running of the interference factor sums Sudakov logarithms associated with the threshold region, but does not bring in any infrared scale dependence.  Understanding it more thoroughly thus appears a promising route to understanding the transition to the elastic limit of hadron-hadron scattering.

\section*{Acknowledgments}

O.Z.L. thanks Jiunn-Wei Chen and National Taiwan University for hosting while some of this work was completed. The work of  S.F. and O.Z.L. was supported by DOE Award No. DE-FG02-04ER41338.

\begin{appendix}

\section{DIS final jet function to $\mathcal{O}(\alpha_s)$ with delta regulator\label{app:jetfndeltareg}}

In this section, we calculate the DIS jet function with the delta regulator. The final jet function is defined in Eq.~\eqref{eq:06} and has been previously calculated to $\mathcal{O}(\alpha_s)$ in Refs.~\cite{Manohar:2003vb,Chay:2013zya,Hornig2009,Bauer2010,Becher2006a} with different regulators. Here we use the delta-regulator prescription introduced in Ref.~\cite{Chiu:2009yx} with $m_g^2$ in the gluon propagator and two delta regulators for two Wilson lines. The delta regulators are added to the collinear and soft Wilson lines the same way as in Sec.~IV. The $\mathcal{O}(\alpha_s)$ Feynman diagrams for the DIS jet function are shown in Fig.~4 where we omit the mirror images of Figs.~4(a) and 4(b).

\begin{figure}[h!]
\centering
\includegraphics[width=0.7\textwidth]{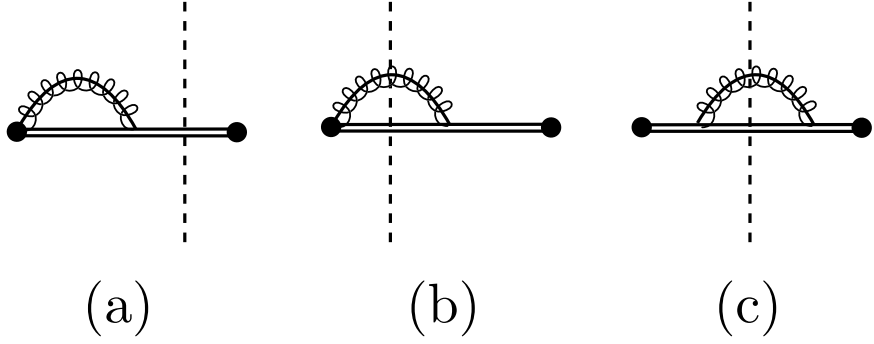}
\caption{$\mathcal{O}(\alpha_s)$ Feynman diagrams for the $\bar n$ jet function.}
\end{figure}

The naive amplitude for virtual gluon emission in Fig.\,4(a) is
\begin{align}\label{eq:A1}
\hat M_a^{\text{jet}}=&
(2ig^2C_F)\mu^{2\epsilon}\delta(r)\int\frac{d^Dq}{(2\pi)^D}
\frac{n\cdot(p-q)}{q^2-m_g^2+i\epsilon}\oneov{(p-q)^2-\Delta_1+i\epsilon}\oneov{n\cdot q+\delta_2+i\epsilon}\,,
\end{align}
where $p_X$ is the DIS final jet momentum. We let $p_X^\mu=(p_X^+,p_x^-,p_{X\perp})=(Q,r,0)$ and \req{eq:A1} becomes
\begin{align}
\hat M_a^{\text{jet}}&=\paren{-\frac{\alpha_sC_F}{2\pi}}\delta(r)\bigg\{-\oneov{\epsilon}\paren{\ln\frac{\delta_2}{p^+}+1}-\ln\frac{\mu^2}{m_g^2}\paren{\ln\frac{\delta_2}{p^+}+1}\nonumber\\
&-\left[\ln\frac{\Delta_1}{m_g^2} \ln\paren{1-\frac{\Delta_1}{m_g^2}}+1-\frac{\Delta_1/m_g^2}{\frac{\Delta_1}{m_g^2}-1}\ln\frac{\Delta_1}{m_g^2}+Li_2\paren{\frac{\Delta_1}{m_g^2}}-\frac{\pi^2}{6}\right]\bigg\}\,,\label{eq:A2}
\end{align}
which has the same form as the naive amplitude of the DIS $n$-collinear function Fig.~2(a). The zero bin for Fig.~4(a) is
\begin{align}
\hat M_{a\phi}^{\text{jet}}&=
(2ig^2C_F)\mu^{2\epsilon}\delta(r)\int\!\frac{d^Dq}{(2\pi)^D}\oneov{q^+}\oneov{\paren{q^--\frac{q_\perp^2+m_g^2-i\epsilon}{q^+}}}\oneov{q^-+\delta_1-i\epsilon}\oneov{-q^+-\delta_2-i\epsilon}\nonumber \\
&=\paren{-\frac{\alpha_s C_F}{2\pi}}\delta(r)\bigg\{ \oneov{\epsilon^2}+\oneov{\epsilon}\ln\frac{\mu^2}{\delta_1\delta_2}+\ln\paren{\frac{\mu^2}{m_g^2}}\ln\paren{\frac{\mu^2}{\delta_1\delta_2}}
\nonumber \\&
-\oneov{2}\ln^2\paren{\frac{\mu^2}{m_g^2}}-Li_2\paren{1-\frac{\delta_1\delta_2}{m_g^2}}+\frac{\pi^2}{12}\bigg\}\,,\label{eq:A3}
\end{align}
which, as expected, has the same form as the zero-bin amplitude of DIS $n$-collinear function Fig.~2(a).
Including the mirror image diagram, the amplitude of final jet function for virtual gluon emission is
\begin{align}
M_a^{\text{jet}}&=2(\hat M_a^{\text{jet}}-\hat M_{a\phi}^{\text{jet}})
\nonumber \\
&=\paren{-\frac{\alpha_sC_F}{2\pi}}2\delta(r)\bigg\{-\oneov{\epsilon^2}-\oneov{\epsilon}\ln\frac{\mu^2}{\Delta_1}-\oneov{\epsilon}-\ln\frac{\mu^2}{m_g^2}\ln\frac{\mu^2}{\Delta_1}-\ln\frac{\mu^2}{m_g^2}-\ln\frac{\Delta_1}{m_g^2}\ln\paren{1-\frac{\Delta_1}{m_g^2}}
\nonumber\\&
+1-\frac{\Delta_1/m_g^2}{\frac{\Delta_1}{m_g^2}-1}\ln\frac{\Delta_1}{m_g^2}-\frac{\pi^2}{12}
+Li_2\paren{\frac{\Delta_1}{m_g^2}}+\oneov{2}\ln^2\!\paren{\frac{\mu^2}{m_g^2}}-Li_2\!\paren{1-\frac{\delta_1\delta_2}{m_g^2}}\bigg\}\,.\label{eq:A4}
\end{align}

The naive amplitude for the real gluon emission in Fig.~4(b) is
\begin{align}
\hat M_b^{\text{jet}}&=
(4\pi g^2C_F)\frac{\mu^{2\epsilon}\,n\cdot p_X}{p_X^2-\Delta_1+i\epsilon}\int\! \frac{d^Dq}{(2\pi)^D}\frac{n\cdot(p-q)}{n\cdot q+\delta_2}
\delta(q^2\!-\!m_g^2)\delta[(p-q)^2-\Delta_1]\theta(p^+\!-q^+)\theta(p^-\!-q^-)\label{eq:A5}
\end{align}
where we use $\Delta_1$ to regulate $\bar n$-direction final jets.  Carrying out the integral, we have
\begin{equation}\label{eq:A6}
\hat M_b^{\text{jet}}=\paren{\frac{\alpha_sC_F}{2\pi Q}}\left\{\delta(z)\ln(-\frac{\delta_1}{Q})\ln\paren{\frac{\delta_2}{p^+}}+\delta(z)\ln(-\frac{\delta_1}{Q})-\paren{\oneov{z}}_+\ln\frac{\delta_2}{p^+}-\paren{\oneov{z}}_+\right\}\,.
\end{equation}
The zero bin for this amplitude is,
\begin{align}
\hat M_{b\phi}^{\text{jet}}&=(-4\pi g^2C_F)\mu^{2\epsilon}\frac{n\cdot p_X}{p_X^2-\Delta_1}\int\frac{d^Dq}{(2\pi)^D}\oneov{(q\cdot n+\delta_2)}
\delta(q^2-m_g^2)\theta(p^--q^-)\delta\paren{p^--q^--\frac{\Delta_1}{p^+}} 
\nonumber\\&
=\paren{-\frac{\alpha_sC_F}{2\pi Q}}\bigg\{\oneov{\epsilon}\left[-\delta(z)\ln(-\frac{\delta_1}{Q})+\paren{\oneov{z}}_+\right]\nonumber \\
&+\ln\paren{\frac{\delta_2 Q}{\mu^2}}\left[-\delta(z)\ln(-\frac{\delta_1}{Q})+\paren{\oneov{z}}_+\right]+\delta(z)Li_2\paren{\frac{Q}{\delta_1}}+\paren{\frac{\ln z}{z}}_+\bigg\}\,.\label{eq:A8}
\end{align}
Including the mirror image diagram, the amplitude of final jet function for real gluon emission in Fig.~4(b) is
\begin{align}
M_b^{\text{jet}}&=2(\hat M_b^{\text{jet}}-\hat M_{b\phi}^{\text{jet}})
\nonumber \\ &
=\frac{\alpha_sC_F}{2\pi Q}2\bigg\{\paren{\delta(z)\ln(-\frac{\delta_1}{Q})-\oneov{z_+}}\paren{1+\ln \frac{\mu^2}{n\cdot p Q}}\nonumber \\
&+\oneov{\epsilon}\left[-\delta(z)\ln(-\frac{\delta_1}{Q})+\paren{\oneov{z}}_+\right]+\delta(z)\left[\oneov{3}\pi^2-\oneov{2}\ln^2\frac{\delta_1}{Q}+\pi i\ln\frac{\delta_1}{Q}\right]+\paren{\frac{\ln z}{z}}_+\bigg\}\,.\label{eq:A9}
\end{align}
The naive amplitude for real gluon emission in Fig.~4(c) is
\begin{align}
\hat M_c^{\text{jet}}&=\paren{\oneov{2\pi}}(-g^2C_F)\frac{(in\cdot p_X)^2}{(p_X^2-\Delta_1)^2}(D-2)\mu^{2\epsilon}\int\frac{d^Dq}{(2\pi)^D}(i)(-2\pi i)\delta(q^2-m_g^2)\nonumber\\
&\times(-2\pi i)\delta[(p-q)^2-\Delta_1]\cdot in\cdot(p-q)\frac{q_\perp^2}{[n\cdot(p-q)]^2}\theta(p^+-q^+)\theta(p^--q^-) \notag \\
&=\paren{\frac{\alpha_sC_F}{2\pi Q}}\oneov{2}\left[-\delta(z)\ln\frac{\delta_1}{Q}+\paren{\oneov{z}}_+\right]\,.\label{eq:A11}
\end{align}
The zero bin for this diagram is
\begin{align}
\hat M_{c\phi}^{\text{jet}}&=\paren{\oneov{2\pi}}(-g^2C_F)\paren{\frac{(in\cdot p_X)^2}{(p_X-\Delta_1)^2}}(D-2)\mu^{2\epsilon}\int\frac{d^Dq}{(2\pi)^D}i(-2\pi i)\delta(q^2-m_g^2)\nonumber \\
&\times\frac{q_\perp^2}{(n\cdot p)^2}(in\cdot p)(-2\pi i)\delta(p^-(p^+-q^+)-\Delta_1)\theta(p^--q^-)\nonumber \\
&=0 \,.\label{eq:A12}
\end{align}
There is no mirror image for Fig.~4(c), so
\begin{equation}\label{eq:A13}
M_c^{\text{jet}}=\hat M_c^{\text{jet}}\,.
\end{equation}
The wave-function contribution to the final jet function is
\begin{equation}\label{eq:A14}
M_{\text{wave}}^{\text{jet}}=\frac{\alpha_sC_F}{2\pi Q}\paren{\oneov{\epsilon}-1-\ln\frac{m_g^2}{\mu^2}}\delta(z)\,.
\end{equation}
Combining all the results above, the $\mathcal{O}(\alpha_s)$ expression for the final jet function up to $\mathcal{O}(\alpha_s)$ is
\begin{align}
M^{\text{jet}}&=M_a^{\text{jet}}+M_b^{\text{jet}}+M_c^{\text{jet}}-\oneov{2}M_{\text{wave}}^{\text{jet}}\nonumber \\
&=\paren{\frac{\alpha_sC_F}{\pi Q}}\bigg\{\oneov{\epsilon^2}+\left[\frac{3}{4}\oneov{\epsilon}\delta(z)+\oneov{\epsilon}\ln\frac{\mu^2}{\Delta_1}\delta(z)+\oneov{\epsilon}\ln(-\frac{\delta_1}{Q})-\oneov{\epsilon}\paren{\oneov{z}}_+\right]\nonumber \\
&+\left[\frac34\paren{\oneov{z}}_++\paren{\frac{\ln z}{z}}_++\ln\paren{\frac{n\cdot p Q}{\mu^2}}\paren{\oneov{z}}_+\right]
+\delta(z)\Bigg[\frac{3}{4}\ln\frac{\mu^2}{m_g^2}+\frac{3}{4}\ln(-\frac{\delta_1}{Q})\nonumber \\
&+\ln\paren{1-\frac{\Delta_1}{m_g^2}}\ln\frac{\Delta_1}{m_g^2}+\frac{7}{8}-\frac{\Delta_1/m_g^2}{\frac{\Delta_1}{m_g^2}-1}\ln\frac{\Delta_1}{m_g^2}+\frac{\pi^2}{4}\Bigg]\bigg\}
\end{align}
This result is independent of $\delta_2$ for the same reason the $n$-collinear function in Eq.~\eqref{eq:72} is independent of $\delta_1$.
The counterterm for the final jet function is
\begin{equation}\label{eq:A16}
\mathcal{Z}_{\bar n}^{\text{jet}}=\delta(z)+\frac{\alpha_sc_F}{2\pi}\left(\delta(z)\left(\frac34+\ln\frac{\mu^2}{\Delta_1}+\ln\left(-\frac{\Delta_1}{(n\cdot p)^2}\right)\right)-\left(\frac{1}{z}\right)_+\right)
\end{equation}
With the choice of $-\delta_1 Q=m_g^2$, we have the renormalized jet function,
\begin{align}
M_{\text{jet}}^R&=\left(\frac{\alpha_sc_F}{\pi}\right)\frac{1}{Q}\bigg\{\frac34\left(\frac{1}{z}\right)_+ +\left(\frac{\ln z}{z}\right)_+ +\ln\frac{Q^2}{\mu^2}\left(\frac{1}{z}\right)_+\nn\\
&+\delta(z)\left[\frac34\ln\left(-\frac{\mu^2}{Q^2}\right)+\frac78+\frac{\pi^2}{4}\right]\,. \label{eq:A17}
\end{align}

\section{KINEMATIC CONSTRAINTS OF THE ZERO-BIN SUBTRACTION WITH THE RAPIDITY REGULATOR}\label{app:zerobin}

The gauge-invariant rapidity regulator automatically ensures the zero bins of the following forms of integrals are scaleless:
\begin{enumerate}
\item the integrals in virtual diagrams;
\item the integrals in real diagrams with measurement functions only involving $\vec k_\perp$\,.
\end{enumerate}
However, in this paper, we encounter integrals for both DIS and DY real soft functions that are not included in the above cases. As a result, we must examine the zero-bin subtraction prescriptions for each of these soft functions carefully to determine whether or not any momenta run into the collinear region.

\begin{figure}[ht!]
\includegraphics[scale=1.0]{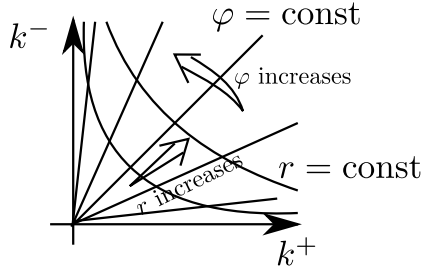}
\centering
\caption{The integration area of $k^+,k^-$ and $r,\varphi$. \label{fig:rapidityvars}}
\end{figure}

After integrating over the perpendicular momentum, the real soft functions for DIS and DY have the following forms respectively:
\begin{align}
I^{\text{DY}}&=\int_0^\infty dk^+\int_0^\infty dk^- \frac{|k^+k^--m_g^2|^{-\epsilon}}{k^+k^-}\theta(k^+k^--m_g^2)|k^+-k^-|^{-\eta}\delta(l-k^+-k^-) \label{eq:B1}\\
I^{\text{DIS}}&=\int_0^\infty dk^+\int_0^\infty dk^- \frac{|k^+k^--m_g^2|^{-\epsilon}}{k^+k^-}\theta(k^+k^--m_g^2)|k^+-k^-|^{-\eta}\delta(l-k^-)\label{eq:B2}
\end{align}
In order to illustrate the origins of the rapidity divergences and the zero bins, we choose a different set of the variables,
\begin{equation}\label{eq:B3}
k^+=re^\varphi, \quad k^-=re^{-\varphi} 
\end{equation}
so that 
\begin{align}
I^{\text{DY}}&=\int_{-\infty}^\infty d\varphi \int_{m_g}^\infty dr\frac{2^{1-\eta}}{r^{1+\eta}}\frac{|r^2-m_g^2|^{-\epsilon}}{|\sinh\varphi|^\eta}\frac{1}{\cosh\varphi}\delta\left(r-\frac{l}{\cosh\varphi}\right)\label{eq:B4}\\
I^{\text{DIS}}&=\int_{-\infty}^\infty d\varphi\int_{m_g}^\infty dr\frac{2^{1-\eta}}{r^{1+\eta}}\frac{|r^2-m_g^2|^{-\epsilon}}{|\sinh\varphi|^\eta}\delta\left(r-le^\varphi\right)e^\varphi\,. \label{eq:B5}
\end{align}
As we can see in Eqs.\,\eqref{eq:B1} and \eqref{eq:B2}, $|k^+-k^-|\to\infty$ can bring in both a rapidity divergence and an ultraviolet divergence. We separate these two types of divergences by working with the $r$ and $\varphi$ variables, because the rapidity divergence is only brought in by $|\sinh\varphi|\to\infty$, and the infrared regulator $m_g^2$ distinguishes an infrared divergence from a rapidity divergence. We illustrate the relations of these two sets of variables in Fig.\,\ref{fig:rapidityvars}. The hyperbolas show the on-shell
condition $k^+k^-=k_\perp^2+m_g^2$, and the zero bins are the rapidity regions $k^+\gg k^-$, $k^-\gg k^+$, or $\varphi\gg 0$, which is also known as the collinear contribution to the soft function.

\begin{figure}[ht!]
\includegraphics[scale=0.8]{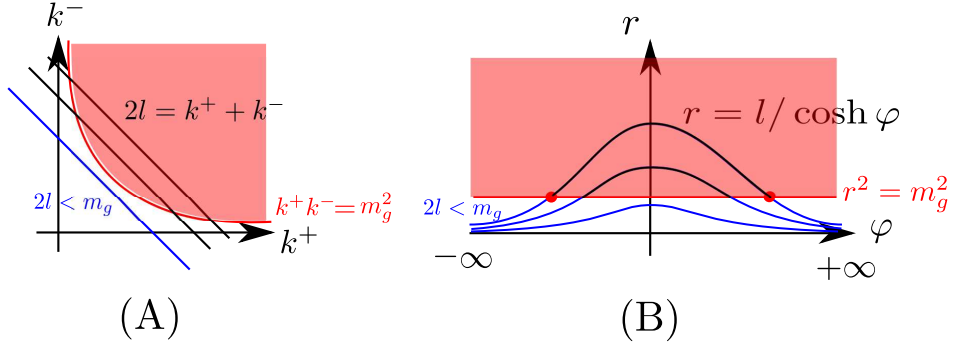}
\centering
\caption{The kinematic constraints for the DY real soft function. (A) is the kinematic constraint in $(k^+,k^-)$ space; (B) is the kinematic constraint in $(r,\varphi)$ space.}\label{fig:DYconstraint}
\end{figure}

\begin{figure}[ht!]
\includegraphics[scale=1.0]{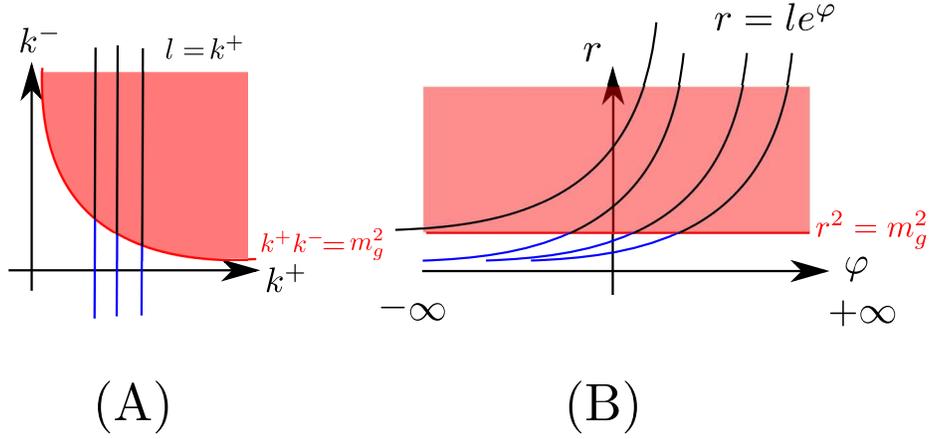}
\centering
\caption{The kinematic constraints for the DIS real soft function. (A) is the kinematic constraint in $(k^+,k^-)$ space; (B) is the kinematic constraint in $(r,\varphi)$ space.}\label{fig:DISconstraint}
\end{figure}

The kinematic constraints are shown in Figs.~\ref{fig:DYconstraint} and \ref{fig:DISconstraint}.
In Fig.\,\ref{fig:DYconstraint}, the (red) shaded part is the integration area, which is constrained by the infrared regulator $m_g^2$. The black lines are the constraints brought in by the measurement function. In Fig.\,\ref{fig:DYconstraint}(A), while $l$ becomes large, it is difficult to tell whether the zero bins $k^+\gg k^-$ or $k^-\gg k^+$ contribute to the naive soft function integral. However, in Fig.\,\ref{fig:DYconstraint}(B), it is very clear that when integrating over the black curve $r\cosh\varphi=l$, $r^2=m_g^2$ cuts off all the collinear contributions from $\varphi\to +\infty$ or $\varphi\to-\infty$.

Therefore, we can conclude that there is no rapidity divergence in the DY real soft function. Interestingly because of the constraint from the measurement function, $r$ is always bounded by $l$, which suggests that we do not have the ultraviolet divergence for this function either.

We analyze the DIS real function in a similar manner in Fig.~\ref{fig:DISconstraint}.  Because the infrared regulator  does not exclude the region $\varphi\to\infty$, collinear momenta contribute to the integral, which brings in the rapidity divergence and requires the zero-bin subtraction.

Carrying out the integrals for the DY and DIS real soft functions
\begin{align}
I^{\text{DY}}&=\int_{-\arccosh(l/m_g)}^{\arccosh(l/m_g)}\frac{1}{2^\eta l^{1+\eta}}\frac{\cosh^\eta\varphi}{|\sinh\varphi|^\eta}\left|\frac{l^2}{\cosh^2\varphi}-m_g^2\right|^{-\epsilon}d\varphi
\nonumber \\
&=\frac{\Gamma(\epsilon)\Gamma(1-\epsilon)}{2^\eta (m_g^2)^\epsilon}\frac{1}{(l^2-m_g^2)^{\frac{1+\eta}{2}}}-\frac{\Gamma(1-\epsilon)\Gamma((1-\eta)/2)}{2^\eta\epsilon\Gamma\left(\frac{1-2\epsilon-\eta}{2}\right)}\frac{1}{(l^2-m_g^2)^{\frac{1+2\epsilon+\eta}{2}}}+\mathcal{O}\left(\frac{l^2}{a^2}-1\right)^{-3/2}
\label{eq:B7}\\
I^{\text{DIS}}&=\int_{\ln(m_g/l)}^\infty \frac{2^{1-\eta}}{l^{1+\eta}}\frac{e^\varphi}{e^{\varphi(1+\eta)}}\frac{|l^2e^{2\varphi}-m_g^2|^\epsilon}{|\sinh\varphi|^\eta}d\varphi 
\nonumber \\&
=\frac{2^{-\eta}}{l^{1+\eta}(m_g^2)^{\epsilon}}\Gamma(\epsilon)\,.\label{eq:B9}
\end{align}
For DY, Eq.\,\eqref{eq:B7} shows that the $\epsilon$ ultraviolet poles cancel between the two terms, and $\eta$ and $\epsilon$ do not regulate $l$ in the factors $(l^2-m_g^2)^{-(1+\eta)/2}$ and $(l^2-m_g^2)^{-(1+2\epsilon+\eta)/2}$. However for DIS, we can extract both rapidity and ultraviolet poles in Eq.\,\eqref{eq:B9}. This analysis clearly shows that the zero-bin subtraction is required only in the presence of the rapidity divergences.

The kinematic constraints seen in Fig.~\ref{fig:DISconstraint} actually produce two distinct zero-bin subtractions in DIS: the first is the ``intuitive'' collinear area in which $k^-\gg k^+$ with $l$ fixed.  This case corresponds to $\phi\to-\infty$ with $r$ fixed.  The second collinear area occurs when $k^+\gg k^-$, because $l$ is large and the measurement function $\delta(l-k^+)$ fixes $k^+=l$.   In DY, we cannot separate the limits $k^+\gg k^-$ and $k^-\gg k^+$ in the integrand of \req{eq:B7} because this requires letting $l$ become large, which opens up phase space at \emph{both} $\phi$ large and positive and $\phi$ large and negative, see Fig.~\ref{fig:DYconstraint}(B). Therefore DY does not have distinct $k^+\gg k^-$ and $k^-\gg k^+$ areas, which is equivalent to the statement that there is no zero-bin subtraction.  

\end{appendix}

\end{document}